\documentclass[preprint,12pt]{emulateapj}
 
\usepackage{apjfonts}
\usepackage{natbib}
\usepackage{amssymb,amsmath}
\usepackage{graphicx}
\usepackage[colorlinks,urlcolor=red,citecolor=blue,linkcolor=cyan]{hyperref} 

\def\ltsima{$\; \buildrel < \over \sim \;$}
\def\lsim{\lower.5ex\hbox{\ltsima}}
\def\gtsima{$\; \buildrel > \over \sim \;$}
\def\gsim{\lower.5ex\hbox{\gtsima}}

\def\vsini{$v \sin i_{\star}$\,}
\def\kms{km\,s$^{-1}$}

\shortauthors{Albrecht et al.} \shorttitle{Low obliquities in compact multiplanet systems}
\begin{document}

\title{Low stellar obliquities in compact multiplanet systems}

\author{
  Simon Albrecht\altaffilmark{1}, 
  Joshua N.\ Winn\altaffilmark{1},
  Geoffrey W.\ Marcy\altaffilmark{2},\\
  Andrew W.\ Howard\altaffilmark{3},
  Howard Isaacson\altaffilmark{2},
  John A.\ Johnson\altaffilmark{4}
}
 
\altaffiltext{1}{Department of Physics, and Kavli Institute for
  Astrophysics and Space Research, Massachusetts Institute of
  Technology, Cambridge, MA 02139, USA}

\altaffiltext{2}{Department of Astronomy, University of California,
  Berkeley, CA 94720, USA}

\altaffiltext{3}{Institute for Astronomy, University of Hawaii, 2680
  Woodlawn Drive, Honolulu, HI 96822, USA}

\altaffiltext{4}{California Institute of Technology, Department of
  Astrophysics, Division of Geological an Planetary Sciences,
  Pasadena, CA 91125, USA; David \& Lucile Packard
  j`q1  Fellow, Sloan Fellow}

\altaffiltext{$\star$}{The data presented herein were collected with
 the Keck~I telescope at the W.M.\ Keck
  Observatory, which is operated as a scientific partnership among the
  California Institute of Technology, the University of California and
  the National Aeronautics and Space Administration.}

\begin{abstract}
  We measure the sky-projected stellar obliquities ($\lambda$) in the
  multiple-transiting planetary systems KOI-94 and Kepler-25, using
  the Rossiter-McLaughlin effect. In both cases the host stars are
  well-aligned with the orbital planes of the planets. For KOI-94 we
  find $\lambda=-11\pm11^{\circ}$, confirming a recent result by
  Hirano and coworkers. Kepler-25 was a more challenging case because
  the transit depth is unusually small (0.13\%). To obtain the
  obliquity it was necessary to use prior knowledge of the star's
  projected rotation rate, and apply two different analysis methods to
  independent wavelength regions of the spectra. The two methods gave
  consistent results, $\lambda=7\pm8^{\circ}$ and
  $-0.5\pm5.7^{\circ}$.

  There are now a total of five obliquity measurements for host stars
  of systems of multiple transiting planets, all of which are
  consistent with spin-orbit alignment. This alignment is unlikely to
  be the result of tidal interactions, because of the relatively large
  orbital distances and low planetary masses in the systems. In this
  respect the multiplanet host stars differ from hot-Jupiter host
  stars, which commonly have large spin-orbit misalignments whenever
  tidal interactions are weak.  In particular the weak-tide subset of
  hot-Jupiter hosts have obliquities consistent with an isotropic
  distribution ($p=0.6$), but the multiplanet hosts are incompatible
  with such a distribution ($p\sim 10^{-6}$). This suggests that high
  obliquities are confined to hot-Jupiter systems, and provides
  further evidence that hot Jupiter formation involves processes that
  tilt the planetary orbit.

\end{abstract}


\section{Introduction}
\label{sec:intro}

\setcounter{footnote}{0}

Over the last few years, many stars with exoplanets have been found to
have high obliquities, i.e., large angles between the stellar equator
and the planet's orbital plane
\cite[e.g.][]{hebrard2008,winn2009,queloz2010,cameron2010b,moutou2011,johnson2011,albrecht2012}.
However, for practical reasons almost all of the measurements have
been made for stars with hot Jupiters. Systems with smaller planets,
longer-period planets, and multiple planets remain relatively
unexplored.

For the hot Jupiters, \cite{winn2010} and \cite{albrecht2012b} found
evidence that the obliquities of many systems have been affected by
tidal evolution: the systems for which one would expect planet-star
tidal interactions to be strongest are preferentially found to have
low obliquities.  Systems with weaker tides have a more random
obliquity distribution.  This suggests that at the time of hot Jupiter
formation, before tides had any opportunity to act, the orbital planes
were only loosely correlated with the equatorial planes of their host
stars. This in turn provides evidence that whatever ``migration''
process produces hot Jupiters also causes their orbits to be tilted
away from the initial plane of formation, favoring scenarios such as
planet-planet scattering or the Kozai effect over the once dominant
paradigm of gradual inspiral within the protoplanetary disk.

The interpretation of the hot-Jupiter results is not settled, though,
because the possibility remains that high obliquities are a generic
feature of planetary systems, not specific to hot Jupiter migration.
There are several proposed mechanisms for tilting a star relative to
its protoplanetary disk: chaotic star formation
\cite[e.g.][]{bate2010,thies2011}, magnetic star-disk interactions
\citep{lai2011,foucart2011}, torques due to internal gravity waves
\citep{rogers2012}, and torques due to neighboring stars
\cite{batygin2012}. In these scenarios, we should observe high
obliquities not only in hot-Jupiter systems but also in a broader
class of planetary systems.

One may test this idea by measuring stellar obliquities in systems
with multiple transiting planets. In such systems the planets' orbits
are likely to be coplanar, and presumably mark the plane of the
protoplanetary disk out of which the planets originally formed. If
these systems have low stellar obliquities as a rule, then the high
obliquities in hot-Jupiter systems are probably due to planet
migration. If instead the obliquity distribution of
multiple-transiting systems is similar to that of hot-Jupiter systems,
then the obliquities are clues to more general processes in star and
planet formation, and not specific to hot Jupiters.

The first multiple-transiting system for which the projected obliquity
was measured was Kepler-30 \citep{sanchis2012}. The authors took
advantage of {\it Kepler} photometry and the occurrence of star-spots
to measure the projected obliquity. More recently \cite{hirano2012}
measured the projected obliquity in KOI-94 making use of the
Rossiter-McLaughlin (RM) effect, and \cite{chaplin2013} constrained
the obliquities of Kepler-50 and Kepler-65 using asteroseismology. All
of these systems were found to be consistent with good spin-orbit
alignment.

\begin{table}[t]
  \caption{Relative radial-velocity measurements}
  \label{tab:rvs}
  \begin{center}
    \vskip -0.15in
    \smallskip
    \begin{tabular}{l c c c}
      \hline
      \hline
      \noalign{\smallskip}
      System & Time & RV & Unc. \\
         & (BJD$_{\rm TDB}$) &(m~s$^{-1}$) & (m~s$^{-1}$)  \\
      \noalign{\smallskip}
      \hline
      \noalign{\smallskip}
      Kepler-25 &  $2455761.77513$  &  $     -7.07$&$  3.53$ \\
      Kepler-25 &  $2455761.78302$  &  $     -5.47$&$  3.45$ \\
      Kepler-25 &  $2455761.85205$  &  $     -6.80$&$  3.77$ \\
      \multicolumn{4}{c}{\vdots}\\
      \noalign{\smallskip}
      \hline
      \noalign{\smallskip} 
    \end{tabular}  
  \end{center}
\end{table}

In this work we present an obliquity determination for the Kepler-25
multiple-transiting system, as well as an independent observation of
the KOI-94 system. Between these and the previously published
measurements, there are now five multiple-exoplanet systems for which
we have information about the stellar obliquity (and of course the
Solar system provides a sixth multiple-planet system). We are now in a
position to make a statistical comparison between the multiple-planet
systems and the hot-Jupiter systems.

The plan of this paper is as follows. The observations are described
in Section~\ref{sec:obs}. Section~\ref{sec:koi94} presents the results
for KOI-94. Section~\ref{sec:koi244} gives the results for Kepler-25,
which were obtained with two independent analysis methods because of
the relatively challenging nature of the detection. The first method
involved analyzing the ``anomalous radial velocity'' due to the RM
effect (Section~\ref{sec:koi244_rvs}), and the second method involved
direct modeling of the line-profile distortions
(Section~\ref{sec:koi244_shape}). As a test of the latter method, an
analysis of archival spectra of the HAT-P-2 system is also
presented. Finally, Section~\ref{sec:comparison} presents statistical
comparisons of the stellar obliquities in multiple-planet systems and
hot-Jupiter systems.

\section{Observations}
\label{sec:obs}

All the observations analyzed here were obtained with the Keck\,I
telescope and its High Resolution Spectrograph
\citep[HIRES;][]{vogt1994}. We observed KOI-94 during the night of
2012~August 9/10, when it was transited by KOI-94.01. We observed
Kepler-25 on two nights coinciding with transits of the largest planet
c (2011~July~18/19 and 2012~May~31/June~1). We determined relative
radial velocities in the usual way for HIRES, by analyzing the stellar
spectra filtered through an iodine cell. The iodine absorption lines
cover the wavelength range from about 500 to 600~nm. The analysis was
performed with a descendant of the original code by \cite{butler1996}.
The RVs of the Kepler-25 and KOI-94 systems are presented in
Table~\ref{tab:rvs}.

\section{KOI-94}
\label{sec:koi94}

The KOI-94 system has been studied in detail by \cite{weiss2013}. It
harbors four planets in orbit around a late F-star (Table~\ref
{tab:koi94_overview}). KOI-94.01 is the largest of these planets,
blocking nearly $0.8$\% of the starlight during transits. By analyzing
the light-curve of a mutual planet--planet eclipse in front of the
stellar disk, \cite{hirano2012} showed that the mutual inclination
between KOI-94.01 and KOI-94.03 is low ($1.15\pm0.55^{\circ}$).  This
suggests that the planets have not been dynamically disrupted and that
their orbits are faithful indicators of the plane of their formation.

\begin{table}[t]
 \caption{Characteristics of KOI-94}
 \label{tab:koi94_overview}
 \begin{center}
 \vskip -0.25in
 \smallskip
     \begin{tabular}{l l l}
	\hline
	\hline
	\noalign{\smallskip}
        KIC &  6462863    &  \tablenotemark{$\star$} \\ 
	Kepler magnitude  & $12.2$& \tablenotemark{$\star$} \\     
        $T_{\rm eff}$ & $6182 (58)$~K&\tablenotemark{$\dagger$} \\
        $\log g$   & $4.182 (0.066)$&\tablenotemark{$\dagger$} \\
        Metallically, [Fe/H]   & $0.0228(20)$ &\tablenotemark{$\dagger$} \\
        Projected stellar rotation speed, $v \sin i_{\star}$  & $7.3(0.5)$~km\,s$^{-1}$ &\tablenotemark{$\dagger$} \\
        Stellar radius, $R_{\star}$ & $1.52(14)~R_{\odot}$&\tablenotemark{$\dagger$} \\
        Stellar mass, $M_{\star}$ & $1.277(50)~M_{\odot}$&\tablenotemark{$\dagger$} \\
	\noalign{\smallskip}
        \noalign{\smallskip}
	\hline
	\noalign{\smallskip}
        \noalign{\smallskip}  
         \multicolumn{3}{l}{$\star$ Data from MAST archive {\tt  http://archive.stsci.edu/kepler/}}\\
        \multicolumn{3}{l}{$\dagger$ Data from \cite{weiss2013}}\\
     \end{tabular}
\end{center} 
\end{table}

\begin{figure*}
  \begin{center}
    \includegraphics[width=14.5cm]{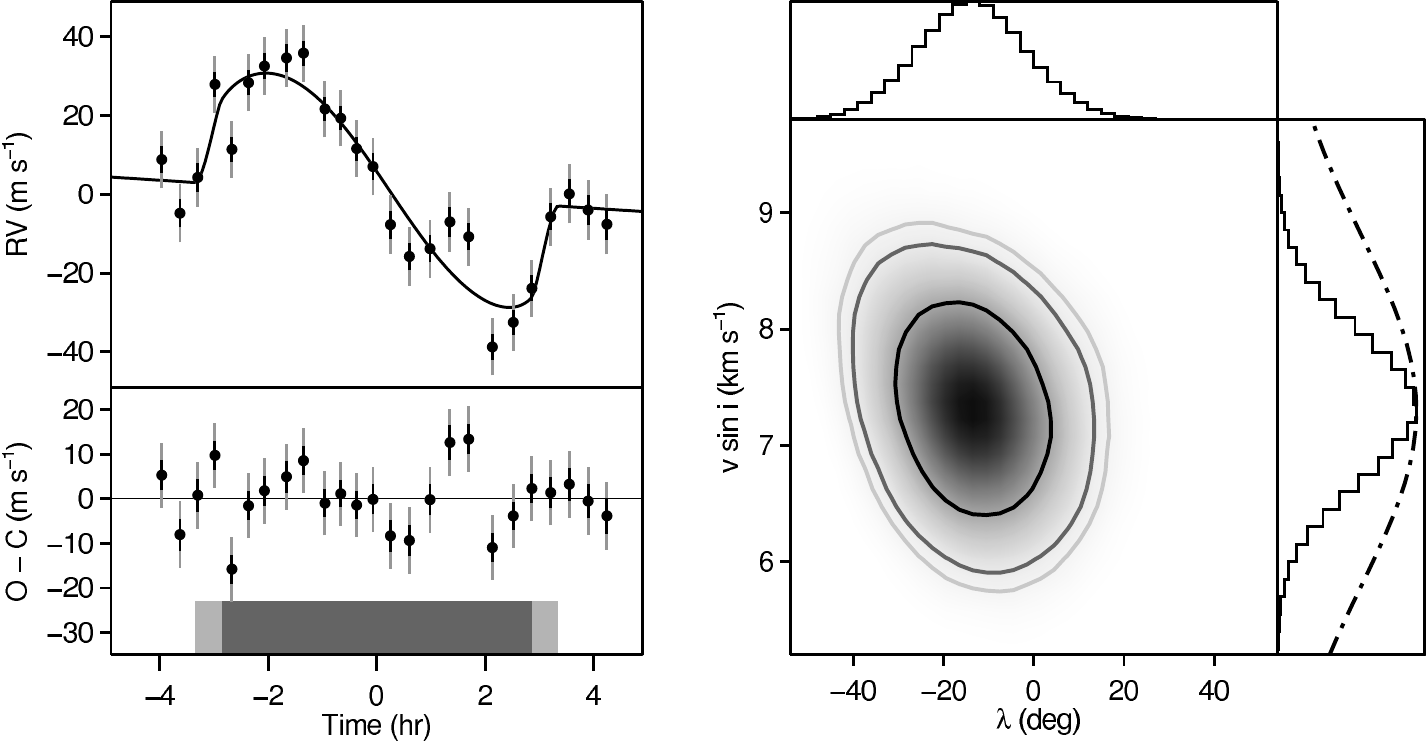}
    \caption {\label{fig:koi94_rv_results} {\bf Rossiter-McLaughlin
        effect for KOI-94.} {\it Left:} Apparent RV variation spanning
      the transit of planet KOI-94.01 on 2012~August~9/10.
      Black error bars show internal
      uncertainties as estimated by the RV measurement routine;
      gray error bars show uncertainties after adding 
      a 'stellar jitter' term in quadrature with the internal uncertainties to obtain a reduced $\chi^2$ of
      $1$. In the upper left panel, the solid curve represents the best-fitting
      model. The lower left panel displays the residuals, with the
      light and dark gray bars indicating
      the calculated times of first, second, third and fourth contact. {\it
        Right panels:} Parameter distributions, based on an MCMC analysis. The main
      plot shows the posterior in the \vsini--$\lambda$ plane, with
      dark shading indicating high probability. The black, dark gray,
      and light gray lines outline the two-dimensional $68.3$\,\%,
      $95$\,\%, and $99.73$\,\% confidence limits. The one-dimensional
      projections of this posterior are shown on the upper and right sides.
      We find $\lambda=-11\pm11^{\circ}$ and
      \vsini$=7.3\pm0.6$\,\kms. The dashed line indicates the prior
      knowledge on \vsini\ ($7.3\pm1.5$\,\kms) which was adopted for
      this analysis.}
  \end{center}
\end{figure*}

By observing the RM effect with Subaru and its High Dispersion
Spectrograph, \cite{hirano2012} measured a low projected obliquity
($\lambda=-6_{+13}^{-11}$$^{\circ}$) for the host star, relative to
the orbit of KOI-94.01. Here, we present an analysis of the very same
transit, based on data obtained with a different telescope. To analyze
the RVs we used the approach of \cite{albrecht2012b}, which is based
on the analytic description of the RM effect by \cite{hirano2011} to
model the RM effect. The model takes into account the rotation of the
star, macroturbulence \citep[e.g.][]{gray2005}, thermal broadening,
and line-broadening due to the finite resolution of the
spectrograph. We added the prescription for the convective blueshift
developed by \citet{shporer2011}, as implemented by
\cite{albrecht2012}. We assumed the convective blueshift to have an
amplitude of $1$~\kms, larger than that of the Sun ($0.5$\,\kms)
because it is thought that slightly hotter stars such as KOI-94 have
stronger convective blueshifts \citep[see][and references
therein]{shporer2011}.

Along with the RVs, we analyzed the transit photometry obtained with
the {\it Kepler} telescope in its short cadence mode (one-minute
sampling). We fitted the photometric data with the \cite{mandel2002}
transit model to determine the orbital period ($P$), the time of a
particular mid-transit ($T_{\rm c}$), and the geometric parameters of
the transit. The geometric parameters are the stellar radius in units
of the orbital distance ($R_{\star}/a$), the planet-to-star radius
ratio ($R_{\rm p}/R_{\star}$), and the cosine of the orbital
inclination ($\cos i_{\rm o}$). We assumed a quadratic limb-darkening
law and allowed both of the coefficients to be free parameters. The
light curve was fitted simultaneously with the RVs. To calculate the
anomalous RV due to the RM effect, we also assumed a quadratic
limb-darkening law, with priors on the coefficients based on the
tabulated values of \cite{claret2000}: $u_{1, {\rm RM}}=0.35$ and $u_{2,
  {\rm RM}}=0.35$. We allowed ($u_{1, {\rm RM}}+u_{2, {\rm RM}}$) to be
adjusted, with prior constraint centered on the tabulated value of
$0.7$ and with a Gaussian width of $0.1$. The difference between the
two parameters ($u_{1, {\rm RM}}-u_{2, {\rm RM}}$) was held fixed at the
tabulated value of $0.0$ since the difference is only weakly
constrained by the data and in turn has little effect on the other
parameters.

Our prior on the projected stellar rotation speed ($v \sin i_{\star}$)
was based on the determination by \cite{weiss2013} using Spectroscopy
Made Easy (SME; see Table~\ref{tab:koi94_overview}). This
spectroscopic modeling code is described by
\cite{valenti1996,valenti2005}. Our confidence interval for this prior
was enlarged to $1.5$~km\,s$^{-1}$, because the \vsini value measured
from the broadening of stellar absorption lines might not be fully
representative for the projected rotation speed of the stellar surface
area covered during transit. For example depending on the impact
parameter, differential rotation might lead to such a mismatch. Our
prior on the macroturbulent velocity is based on
\cite{valenti2005}. From their equation (1) we obtained a
macroturbulence velocity ($\zeta$) of $5.17$~\kms, for which we
assumed a confidence interval of $1.5$~\kms.\footnote{This represents
  a different approach form the one we adopted in \cite{albrecht2012b}
  where we used the relationship from \cite{gray1984} to estimate
  $\zeta$. However adopting a prior on $\zeta$ from \cite{gray1984},
  and at the same time adopting a prior on \vsini derived with the SME
  tool is inconsistent. This is because SME uses its own estimate of
  the $\zeta$ when extracting \vsini from the line width. We 
  tested if using the different priors makes a material difference, which is
  not the case.}

To isolate the RM signal one must subtract (or model simultaneously)
the orbital RV variation. One possibility is to subtract a model of
the orbital RV variation based on the RV semiamplitude ($K_{\star}$)
obtained by \cite{weiss2013}, which was based on RVs obtained
sporadically over several months. We did not choose this approach out
of concern that apparent RV variation due to starspots or other
astrophysical noise can depend strongly on timescale.
Starspot-induced signals, for example, can introduce slow drifts in
the measured RV signal on a particular night. Such a signal would be
averaged out in a data set obtained over many stellar rotation
periods. See \cite{albrecht2012b} for examples of this effect and how
it can influence measurements of $\lambda$. Consequently we did not
apply a prior constraint on $K_{\star}$ or on the systemic velocity
$\gamma$ in our analysis.\footnote{See also \cite{isaacson2010} for a
  discussion of stellar jitter and its influence on RV measurements.}

For the estimation of the uncertainty intervals, we used an MCMC
algorithm \citep[see, e.g.,][]{tegmark2004}. In Table
\ref{tab:koi94_results} we report the results derived from the
posterior, where the quoted uncertainties exclude $15.85$\,\% of all
values at both extremes, and encompass $68.3$\,\% of the total
probability. Figure~\ref{fig:koi94_rv_results} presents the measured
RVs, the best fitting model and the posterior in the \vsini--$\lambda$
plane. The key result is $\lambda=-11\pm11^{\circ}$, a low projected
obliquity between the orbital angular momentum of KOI-94.01 and the
angular momentum of the stellar rotation. \cite{hirano2012} found
$\lambda=-6^{+13}_{-11}$$^{\circ}$, which is consistent with our
results. However, there were some differences in the methods of
analysis. \cite{hirano2012} did not account for correlations between
the uncertainty in $\lambda$ and the uncertainties in $T_c$,
$R_{\star}/a$, $R_{\rm p}/R_{\star}$, $i_{\rm o}$, or $\zeta$; they
did not model the convective blueshift; and they used RV observations
obtained on different nights. For a fairer comparison we repeated the
analysis of their RVs with the same constraints we applied to our own
data. In this way we found $\lambda=-7\pm17^{\circ}$ based on the
Subaru dataset. We therefore conclude that two independent data sets
support the finding of a low stellar obliquity in the KOI-94 system.

\section{Kepler-25}
\label{sec:koi244}

The transiting objects in the Kepler-25 system were confirmed to be
planets by \cite{steffen2012}, through the detection and
interpretation of transit timing variations (TTV). The system was also
recently analyzed by \cite{lithwick2012}, who measured masses of
$7.13\pm2.5$\,$M_{\Earth}$ and $13.1\pm2.6$\,$M_{\Earth}$ for the two
transiting planets b and c. (For comparison the mass of Neptune is
$17.15$\,$M_{\Earth}$.) Table~\ref{tab:koi244_overview} gives the
basic system parameters. Detection of the RM effect for this system is
particularly challenging because the largest planet c blocks only
0.13\% of the starlight. This leads to a low signal-to-noise ratio
(S/N) detection of the RM effect. To gain more confidence in the
results we employed two different techniques for measuring $\lambda$,
relying on two different portions of the stellar spectrum. These two
measurements are largely independent, although for both methods we use
the {\it Kepler} photometry as supporting data. In the first technique
(section~\ref{sec:koi244_rvs}), we analyze the RV time series which is
derived from the iodine region of the spectrum. In the second
technique (section~\ref{sec:koi244_shape}) we do not analyze RVs;
rather, we directly model the deformation of the stellar absorption
lines. For this we use a method which we originally developed for the
analysis of mutual events in eclipsing star systems
\citep{albrecht2007,albrecht2009,albrecht2011,albrecht2012c}. A
similar approach has also been used by
\cite{cameron2010,cameron2010b,miller2010,gandolfi2012} and \citet{brown2012}.

\begin{table}
 \begin{center}
  \caption{Parameters of the KOI-94 system\label{tab:koi94_results}}
    \smallskip 
       \begin{tabular}{l  r@{\,\,$\pm$\,\,}l    }
          \tableline\tableline
          \hline
	  \noalign{\smallskip}
	  Parameter &  \multicolumn{2}{c}{ Values} \\
	  \noalign{\smallskip}	 
	  \hline
	  \noalign{\smallskip}
          \multicolumn{3}{c}{Parameters mainly derived from photometry} \\
          \noalign{\smallskip}
          \hline
	  \noalign{\smallskip}
          Mid-transit time $T_{\rm c}$ [BJD$_{\rm TDB}$$-$2\,400\,000]   &  $54965.74092$ & $0.00014$     \\
          Period, $P$ [days]                                                                  &  $22.342971$  &  $0.000004$   \\
          Cosine orbital inclination KOI-94.01, $\cos i_{\rm o}$             &   $0.0112$    &  $0.0012$ \\   
          Fractional stellar radius,         $R_{\rm \star}/a $                      &   $0.03807$ &  $0.0003$  \\
          Fractional planetary radius,	 $  R_{\rm p}/R_{\star} $              & $0.07019$  & $0.00018$ \\
          $u_{1}+u_{2}$                                                                        &  $ 0.538$ &  $0.018$ \\         
          $u_{1}-u_{2}$                                                                        &  $0.070$ &  $0.053$ \\  
        \noalign{\smallskip}
          \hline
          \noalign{\smallskip}
          \multicolumn{3}{c}{Parameters mainly derived from RVs} \\
          \noalign{\smallskip}
          \hline 
          \noalign{\smallskip}
          Velocity offset, $\gamma$ [m\,s$^{-1}$]                                           & $-2.3$ & $2.1$ \\
          Orbital semi-amplitude, K$_{\star}$  [m\,s$^{-1}$]                 &   $74$ & $64$ \\
          $\sqrt{v \sin i_{\star}} \sin \lambda$  [${\sqrt{\rm km\,s^{-1}}}$]                 &  $-0.527$ & $0.53$ \\
          $\sqrt{v \sin i_{\star}} \cos \lambda$  [${\sqrt{\rm km\,s^{-1}}}$]               & $2.60$ & $0.14$  \\
          Macro turbulence parameter, $\zeta$  [km\,s$^{-1}$]             &  $5.04$ &  $1.5$ \\      
          $u_{1,{\rm RM}}+u_{2,{\rm RM}}$                                                                        &  $0.65$ &  $0.3$ \\         
          \noalign{\smallskip}
          \hline
          \noalign{\smallskip}
          \multicolumn{3}{c}{Indirectly derived parameters} \\
	  \noalign{\smallskip}
          \hline
          \noalign{\smallskip}
          Orbital inclination KOI-94.01,   $i_{\rm o}$   [$^{\circ}$]      &  $89.36$ &  $0.07$    \\
          Full duration, $T_{14}$  [hr]                              & $6.689$ & $0.008$     \\
          Ingress or egress duration, $T_{12}$  [hr]         & $0.477$ &  $0.009$       \\
          Projected stellar rotation speed, $v \sin i_{\star}$ [km\,s$^{-1}$]     &   $7.3$ & $0.6$            \\
          Projected spin-orbit angle,  $\lambda$    [$^{\circ}$]             &      $-11$ & $11$ \\
	  \noalign{\smallskip}
	  \tableline
           \noalign{\smallskip}
           \noalign{\smallskip}
     \end{tabular}
   \end{center}
\end{table}

\subsection{Analyzing the RVs}
\label{sec:koi244_rvs}

The analyis of the Kepler-25 RV time series was similar to the
analysis of the KOI-94 RV time series. Because for Keopler-25 we had
RV measurements from two different transit nights, separated by nearly
one year, we introduced for each night a different velocity offset
($\gamma$) and a different parameter to fit the out-of-transit
velocity slope ($K_{\star}$). We used a prior on \vsini\ based on an
SME analysis (see Table \ref{tab:koi244_overview}). For $\zeta$ we
used $4.85\pm1.5$~\kms, which was obtained in the same way as for
KOI-94. From the tables of \cite{claret2000} we obtained prior
information on the limb-darkening coefficients ($u_{1,{\rm RM}} =0.33$
and $u_{2,{\rm RM}}=0.36$). As the expected RM signal has an amplitude
of only a few m\,s$^{-1}$, the convective blueshift (CB) might have a
significant influence on the observed RM signal
\citep{shporer2011}. Thus we allowed the CB parameter to vary, with
only a weak prior of $1\pm0.5$~\kms\, instead of keeping it fixed as
we did for KOI-94.

Because the planet shows TTVs we used a slightly different approach
for incorporating the {\it Kepler} photometry into our analysis. First
we examined the {\it Kepler} photometry for the two particular
transits observed with HIRES, and derived midtransit times. From these
we computed the ephemeris $T_{\rm c}=2455762.03086\pm 0.00050$\,BJD
and $P=12\fd7203424 \pm 0\fd00003$~days. Second, we measured
midtransit times for all the {\it Kepler} transits and used these to
create a single, phase-folded, high-S/N transit light curve for
Kepler-25\,c. This light curve was then fitted together with the RVs,
providing tight constraints on the geometric transit parameters.

\begin{figure*}
  \begin{center}
    \includegraphics[width=14.cm]{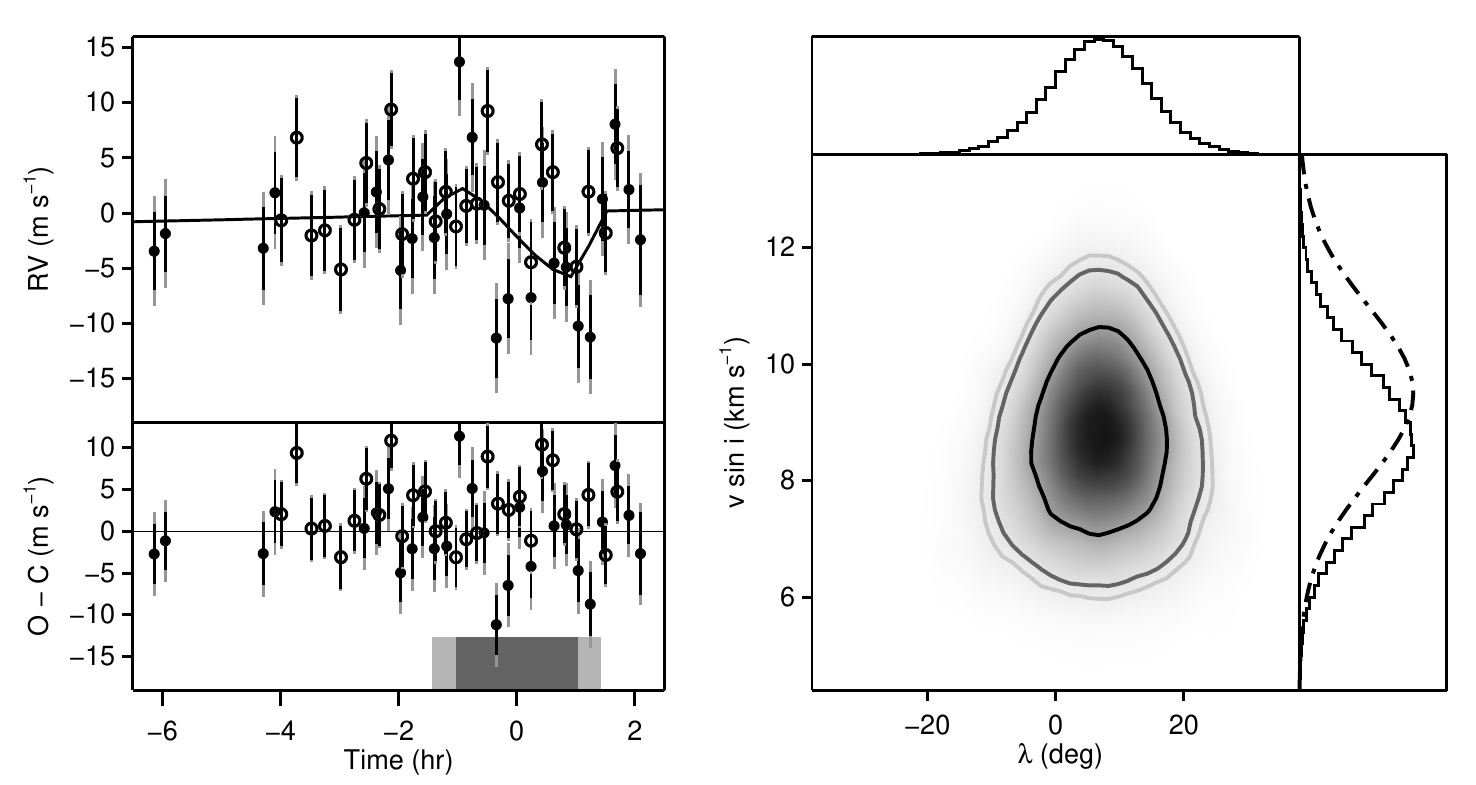}
    \caption {\label{fig:koi244_rv_results} {\bf Measured projected
        obliquity in Kepler-25} The same as
      Figure~\ref{fig:koi94_rv_results} but this time for the
      Kepler-25 system. The RV measurements from the two 
      transit nights are indicated with solid (July 18/19, 2011) and
      open (May/June 31/1, 2012) symbols. We find
      $\lambda=5\pm8^{\circ}$ and \vsini$=8.5\pm0.6$. }
  \end{center}
\end{figure*}

Figure~\ref{fig:koi244_rv_results} shows the RVs from both nights,
and the results for \vsini and $\lambda$ based on an MCMC analysis of
these RVs. The results are given in Table~\ref{tab:koi244_results},
second column. Our measurement of $\lambda=5\pm8^{\circ}$ is consistent
with alignment between the orbital plane of planet c and the stellar
equator.

\begin{table}[t]
 \caption{Kepler-25: Stellar characteristics}
 \label{tab:koi244_overview}
 \begin{center}
 \smallskip
     \begin{tabular}{l l l}
	\hline
	\hline
	\noalign{\smallskip}
        KOI & 244  & \tablenotemark{$\star$} \\ 
        KIC &  4349452  &  \tablenotemark{$\star$} \\ 
	Kepler magnitude  & $10.73$\,mag& \tablenotemark{$\star$} \\     
        $T_{\rm eff}$ & $6301 (82)$K&\tablenotemark{$\dagger$} \\
        $\log g$   & $4.02 (0.1)$&\tablenotemark{$\dagger$} \\
        Metallically, [Fe\/H]   & $-0.10(4)$ &\tablenotemark{$\dagger$} \\
        Projected stellar rotation speed, $v \sin i_{\star}$  km\,s$^{-1}$  & $9.5(0.5)$ &\tablenotemark{$\dagger$} \\
        Stellar radius, $R_{\star}$ & $1.36(13) R_{\odot}$&\tablenotemark{$\ddagger$} \\
        Stellar mass, $M_{\star}$ & $1.22(6) M_{\odot}$&\tablenotemark{$\ddagger$} \\
	\noalign{\smallskip}
        \noalign{\smallskip}
	\hline
	\noalign{\smallskip}
        \noalign{\smallskip}  
         \multicolumn{3}{l}{$\star$ Data from MAST archive {\tt  http://archive.stsci.edu/kepler/}}\\
        \multicolumn{3}{l}{$\dagger$ Obtained using the SME package \cite{valenti1996}}\\
        \multicolumn{3}{l}{$\ddagger$ Data from \cite{steffen2012} }\\
     \end{tabular}
\end{center} 
\end{table}

\begin{figure}
  \begin{center}
    \includegraphics[width=8.cm]{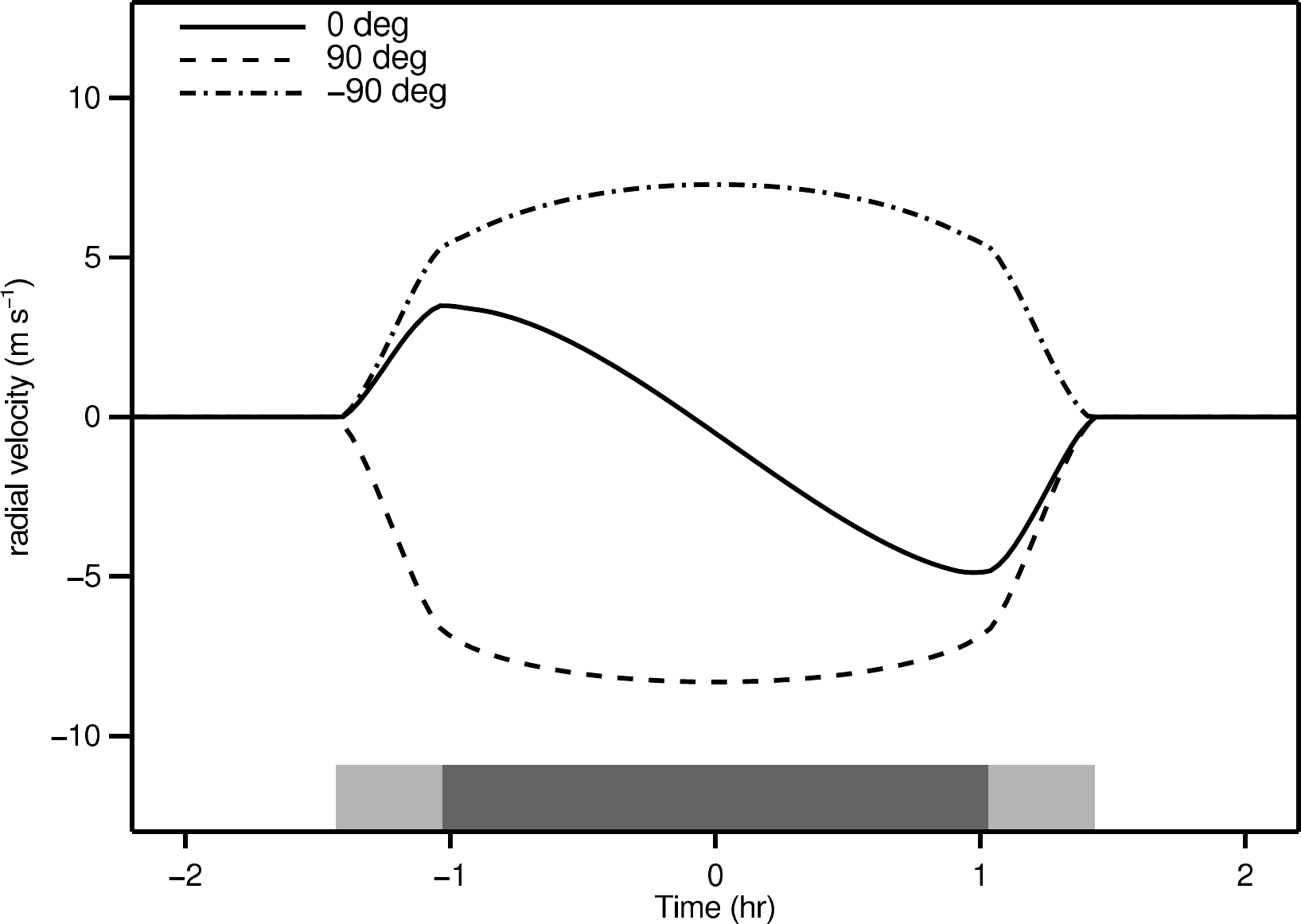}
    \caption {\label{fig:koi244_show} {\bf RM signal for different
        obliquities. } The plot shows the expected RM signal for a system like
      Kepler-25  during transit of planet c. The solid line shows the
      signal for  the parameters printed in the second column
      of Table~\ref{tab:koi244_results} but now with
      $\lambda=0^{\circ}$. The dashed and dashed dotted lines show the
      signals for $\lambda=90^{\circ}$ and $\lambda=-90^{\circ}$. The
      maximum amplitude of the aligned signal is smaller than for the
      strongly misaligned cases. }
  \end{center}
\end{figure}

Because the amplitude of the RM effect is $\approx4$~m\,s$^{-1}$,
comparable to typical uncertainty of a single RV measurement, we must
be skeptical and examine this result further. Why does our algorithm
find such a small uncertainty interval for $\lambda$, given that the
RM signal is so difficult to discern in the time series of the RV
measurements (Figure~\ref{fig:koi244_rv_results})?

We have a great deal of prior knowledge of all the system parameters
relevant for the RM effect, except for $\lambda$, allowing us to
predict accurately the expected characteristics of the RM signal as a
function of $\lambda$. To first order the amplitude of the RM effect
is proportional to the covered surface area and the projected rotation
speed. \citep[See ][for a more detailed
discussion.]{gaudi2007,albrecht2011b} However the amplitude of the RM
effect also depends on $\lambda$ itself.  The amplitude of the RM
signal is nearly twice as large for $\lambda=\pm90^{\circ}$ as it is
for $\lambda$ near $0^{\circ}$ or $180^{\circ}$
(Figure~\ref{fig:koi244_show}). Because of this and because there is a
hint of a prograde signal in the RVs
(Figure~\ref{fig:koi244_rv_results}) the low projected obliquity is
favored.

\begin{figure}
  \begin{center}
    \includegraphics[width=8.cm]{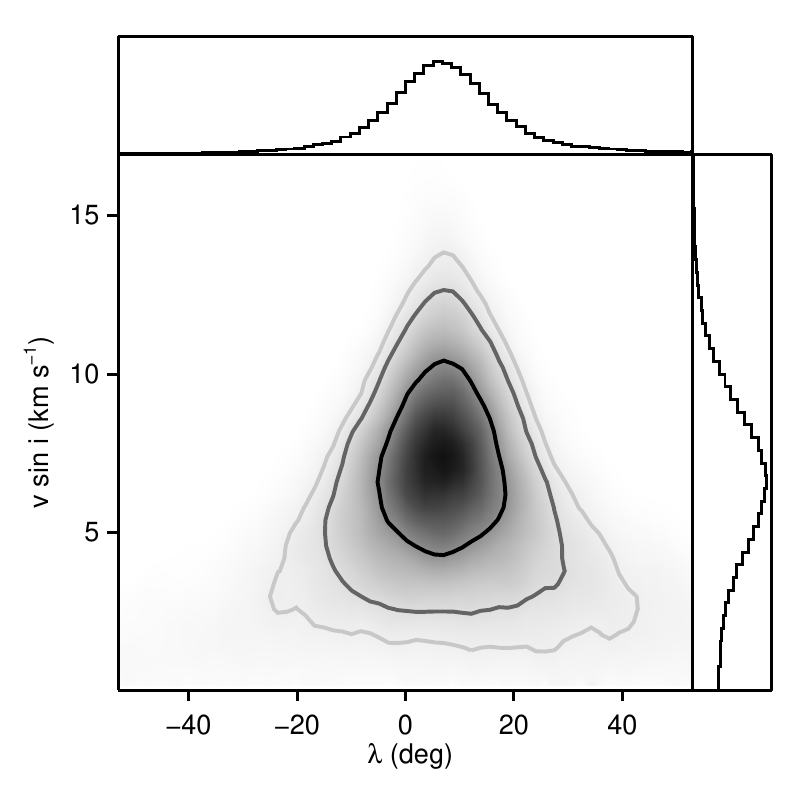}
    \caption {\label{fig:koi244_no_vsini_prior} {\bf Kepler-25 results
        without using prior knowledge on \vsini.} The same as the
      left panels from Figure~\ref{fig:koi244_rv_results} but this
      time without using the \vsini prior. Constequently the
      uncertainties for $\lambda$ and \vsini increased to
      $\lambda=7\pm13^{\circ}$ and \vsini$=6.2\pm3$.}
  \end{center}
\end{figure}

In more detail: we know from the {\it Kepler} photometry that planet c
has a high impact parameter, i.e., it travels near the stellar limb
throughout the transit. We also know {\it a priori} that the star has
a substantial projected rotation speed from the SME analysis.
Combining these two pieces of information we know there is no way to
make the RM effect vanish.\footnote{If the projected stellar rotation
  speed were not known, or if the planet had a low impact parameter,
  then it would be possible to reduce the amplitude of the RM signal
  to arbitrarily small values \cite{albrecht2011b}.}
Figure~\ref{fig:koi244_rv_results} illustrates that we did not make a
high-S/N detection of the RM effect.  We might therefore ask which
values of $\lambda$ would lead to the smallest RM amplitude, or an RM
signal which would be easiest to hide by adjusting other parameters in
our model.  We have just noted that the maximum amplitude of the RM
effect is larger for $|\lambda|$ near $90^{\circ}$ than for $\lambda$
near $0^{\circ}$ or $180^{\circ}$ (Figure~\ref{fig:koi244_show}). This
is because limb darkening attenuates the signal by a factor of
$0.6-0.7$ for a star like Kepler-25. The RM signal has maxima when the
planet crosses the stellar limb where limb darkening is strongest, and
the signal is zero near mid-transit where limb darkening is
weakest. For $|\lambda|$ near $90$ degrees the maximum RM signal
occurs nearer to the center of the stellar disk.  In addition, the
parameters $\gamma$ and $K_{\star}$ (the out-of-transit slope) are
most strongly covariant with $\lambda$ when $\lambda$ is near
$0^{\circ}$ or $180^{\circ}$ \citep{albrecht2011b}.  This is because
of the time-antisymmetry of the RM signal in such cases. In contrast,
for $\lambda$ near $\pm 90^{\circ}$ the RM signal is time-symmetric.
Finally, differential rotation would also weaken the RM signal for
$\lambda = 0^{\circ}$ or $180^{\circ}$ compared to the $\lambda
=\pm90^{\circ}$ signal, though this is a comparatively minor effect
($\sim10$\,\%). Together these factors make it easier to hide an RM
signal with $\lambda = 0$ (or $180$) than $\lambda = \pm
90^{\circ}$. Therefore, it is possible to infer that $\lambda$ must be
near $0^{\circ}$ or $180^{\circ}$ with a sufficiently constraining
upper limit on the amplitude of the RM effect. Here the data prefer
$0^{\circ}$ over $180^{\circ}$ as there is a hint of a prograde RM
signal in the data.  As mentioned above there is only one parameter
that is not at least partly constrained by photometry or prior
knowledge, which is the projected obliquity. This is the qualitative
explanation for the MCMC result of $\lambda = 5\pm8$ degrees. Note
that the {\it a priori} knowledge on \vsini is crucial in this
situation \cite[e.g][]{albrecht2011b}. We illustrate this point by
runing a chain without using the prior knowledge on \vsini, where the
uncertainty interval is enlarged to  $\lambda=7\pm13^{\circ}$. The result
is shown in Figure~\ref{fig:koi244_no_vsini_prior}. 

However it is not completely satisfactory to argue for a low obliquity
based on the absence of a clear signal. Therefore we sought an
independent method to detect the RM effect.  The next section
describes our analysis of a wavelength region in which no iodine lines
are present, and which therefore was not used in the determination of
the RVs. We used this spectral range to make a second independent
measurement of $\lambda$.

\subsection{Measurement of the planet's Doppler shadow}
\label{sec:shape}

In this method of analysis we did not use RVs as proxies for the
distortions in the stellar absorption lines. Instead we analyzed the
line shapes directly to infer $\lambda$. The transiting planet
selectively distorts the line profile, blocking a certain range of
velocity components that ordinarily contribute to the overall line
broadening, a phenomenon referred to as the ``Doppler shadow'' by
\cite{cameron2010}. To model this phenomenon we used a code developed
for double-lined eclipsing binaries
\citep[e.g.][]{albrecht2007,albrecht2012c}. For binary star systems,
the difficulty lays in the additional set of stellar absorption lines
originating from the eclipsing foreground star. In the case of
planetary transits, this particular difficulty is eliminated, as the
planet's emitted light is negligible; rather, the challenge stems from
the small amplitude of the RM signal. In the case of a Jupiter-sized
planet such as HAT-P-2b, about 1\% of the light is blocked from
view. In the case of the transit of planet c in front of Kepler-25
only $0.13$~\% of the light is blocked. For this reason we will
present the application of our code to the Jupiter-sized planet in the
bright ($m_{\rm V}=8.7$) HAT-P-2 system, as a test case. We then
proceed to the more challenging Kepler-25 system.

\subsubsection{The method} We used a two-step process to measure
$\lambda$ from stellar spectra. In the first step we combined the
signals from many stellar absorption lines into one high-S/N
absorption line, which we will call the ``kernel.''  The spectrum is
modeled as the convolution of the kernel and a series of
$\delta$-functions at the central wavelengths of the absorption
lines. This was done for each individual spectrum. In the second step,
we analyzed the distortions that are seen in the kernels, which are
caused by the transit of the planet over the rotating photosphere. As
the first step is different from the approach used by
\cite{albrecht2007}, we discuss it in detail in the following
subsection.\footnote{Previously, \cite{albrecht2007} used the
  broadening function developed by \cite{rucinski1999}. For
  comparison, we have reanalyzed those data using the algorithm
  presented here, finding equivalent results for $\lambda$ and the
  other system parameters. However, for Kepler-25 we found it easier
  to create a template spectrum using the new approach presented here,
  because of the availability of higher-S/N spectra. The HIRES spectra
  of Kepler-25 have a S/N between 50 and 100 in the wavelength range
  $398$ to $479$~nm.}

\begin{table*}
 \begin{center}
   \caption{Results for the Kepler-25 system.\label{tab:koi244_results}}
    \smallskip 
         \begin{tabular}{l  r@{\,\,$\pm$\,\,}l r@{\,\,$\pm$\,\,}l   }
           \tableline\tableline
           \hline
           \noalign{\smallskip}
           Parameter &  \multicolumn{2}{c}{ Values RV} &\multicolumn{2}{c}{ Values Distortion}  \\
           \noalign{\smallskip}	 
           \hline
           \noalign{\smallskip}
           \multicolumn{5}{c}{Parameters mainly derived from photometry} \\
           \noalign{\smallskip}
           \hline
           \noalign{\smallskip}
           Mid-transit time $T_{\rm c}$ [BJD$_{\rm TDB}$$-$2\,400\,000] & $55762.0309$&$0.0005^{\rm a}$  &    $55762.0308$&$0.0005^{\rm a}$   \\
           Period, $P$ [days]                                                                  &  $12.72034$&$0.00003^{\rm a}$   &    \multicolumn{2}{c}{12.7203424   (fixed)}   \\
           Cosine orbital inclination Kepler-25\,c, $\cos i_{\rm o}$           &  $0.0472$&$0.0008$ &    $0.0476$&$0.0008$   \\   
           Fractional stellar radius,         $R_{\rm \star}/a $                      &   $0.0537$&$0.0007$  &    $0.0540$&$0.0007$   \\
           Fractional planetary radius,	 $  R_{\rm p}/R_{\star} $              & $0.0360$&$0.0006$ &    $0.0362$&$0.0007$   \\
           $u_{1}+u_{2}$                                                                        &  $0.569$&$0.020$ &    $0.560$&$0.019$   \\         
           $u_{1}-u_{2}$                                                                        &  $-0.10$&$0.5$ &     $0.04$&$0.6$  \\  
           \noalign{\smallskip}
           \hline
           \noalign{\smallskip}
           \multicolumn{5}{c}{Parameters mainly derived from spectroscopy} \\
           \noalign{\smallskip}
           \hline 
           \noalign{\smallskip}
           Velocity offset 2011, $\gamma$ [m\,s$^{-1}$]                             & $-3.7$&$1.3$ &    \multicolumn{2}{c}{\,\,\,\,\,\,$^{\rm b}$}  \\
           Velocity offset 2012, $\gamma$ [m\,s$^{-1}$]                             & $2.7$&$0.7$ &    \multicolumn{2}{c}{\,\,\,\,\,\,$^{\rm b}$}  \\
           Orbital semi-amplitude 2011, K$_{\star}$  [m\,s$^{-1}$]                 & $-10$&$23$ &        \multicolumn{2}{c}{\,\,\,\,\,\,$^{\rm b}$} \\
           Orbital semi-amplitude 2012, K$_{\star}$  [m\,s$^{-1}$]                 & $-32$&$25$ &        \multicolumn{2}{c}{\,\,\,\,\,\,$^{\rm b}$} \\
           $\sqrt{v \sin i_{\star}} \sin \lambda$ [${\sqrt{\rm km\,s^{-1}}}$]      & $-0.35$&$0.39$ &      \multicolumn{2}{c}{\,\,\,\,\,\,$^{\rm b}$}   \\
           $\sqrt{v \sin i_{\star}} \cos \lambda$  [${\sqrt{\rm km\,s^{-1}}}$]     & $2.9$&$0.23$  &      \multicolumn{2}{c}{\,\,\,\,\,\,$^{\rm b}$}    \\
           Macro turbulence parameter, $\zeta$  [km\,s$^{-1}$]             &  $4.9$&$1.5$ &    $4.1$&$1.1$   \\      
           Convective blueshift  [km\,s$^{-1}$]                                         &  $-1.0$&$0.5$ &      $-0.8$&$0.07$ \\   
           $u_{1 rm}+u_{2 rm}$                                                                     &  $0.70$&$0.10$ &    $0.65$&$0.09$   \\         
           Point Spread Function, PSF  [km\,s$^{-1}$]                             &  \multicolumn{2}{c}{ 3 (fixed)}  &      $4.2$ & $0.4$  \\   
           \noalign{\smallskip}
           \hline
           \noalign{\smallskip}
           \multicolumn{5}{c}{Indirectly derived parameters} \\
           \noalign{\smallskip}
           \hline
           \noalign{\smallskip}
           Impact parameter  Kepler-25\,c, $b$                                           & $0.879$&$0.004$ &    $0.881$  & $0.004$ \\ 
           Orbital inclination Kepler-25\,c,   $i_{\rm o}$   [$^{\circ}$]             &  $87.30$&$0.05$    &    $87.27$&$0.05$     \\
           Full duration, $T_{14}$  [hr]                                                         & $2.860$&$0.009$     &    $2.861$&$0.009$  \\
           Ingress or egress duration, $T_{12}$  [hr]                                     &  $0.405$& $0.014$   &  $0.410$&$0.015$   \\
           Projected stellar rotation speed, $v \sin i_{\star}$ [km\,s$^{-1}$]   &  $8.7$&$1.3$        &    $8.2$&$0.2$$^{\rm c}$   \\
           Projected spin-orbit angle,  $\lambda$    [$^{\circ}$]                   &  $7$&$8$      &     $-0.5$&$5.7$$^{\rm c}$   \\
           \noalign{\smallskip}
           \tableline
           \noalign{\smallskip}
           \noalign{\smallskip}
           \multicolumn{5}{l}{{\sc Notes} ---}\\
           \multicolumn{5}{l}{$^{\rm a}$ We used the priors
             $P=12\fd7203424 \pm  0\fd00003$\, and $T_{\rm
               c}=2455762.03086\pm 0.00050$\,BJD, }\\
           \multicolumn{5}{l}{   \,\,\, as determined using the  {\it Kepler}
             photometry of the appropriate transits (see section ~\ref{sec:koi244_rvs}).} \\
           \multicolumn{5}{l}{$^{\rm b}$ Was not determined.}\\
           \multicolumn{5}{l}{$^{\rm c}$ Here we step directly in  $v
             \sin i_{\star}$ and $\lambda$  as they are less
             correlated than for RV  measurements.}\\
           \noalign{\smallskip}    
     \end{tabular}
   \end{center}
\end{table*}

\subsubsection{Measuring high S/N ratio stellar rotation kernels}

\paragraph{Preparing the spectra.} The new algorithm works as follows.
First we normalize the spectra using observations of fast rotating
B-stars. Specifically we use polynomials fitted to the same and
adjacent orders in the B-star spectrum to normalize the spectral
orders of our science target. We use adjacent orders to remove the
influence of shallow spectral lines present in the B star spectrum.
All spectra are shifted according to the barycentric correction, and a
correction term derived from the measurements of deep telluric lines
in the red wavelength arm of HIRES. These corrections line up the
spectra with an accuracy of $100-200$~m\,s$^{-1}$ (but see also
below). Next, a high-S/N spectrum is created by averaging over all
out-of-transit observations obtained during the night. Now each
spectrum is compared to this high-S/N spectrum to mark and discard bad
pixels. We also make one final small differential correction in the
normalization of all spectra. For this we compare all spectra to the
high-S/N spectrum and fit a third-order polynomial to the residuals,
in which no absorption lines are present. Such a polynomial is created
for each order and spectrum and is used for the correction.

\paragraph{Creating and refining a template.} To combine all the
information contained in the different absorption lines, we need a
sharp-lined template spectrum matching that of the target star. Our
approach to obtain such a template---which matches the observed
spectrum after convolution with a ``master'' absorption line
profile---was inspired by \cite{reiners2003}, but see also
\cite{donati1997} and \cite{rucinski1999} for similar approaches to
obtain high-S/N kernels.

It is not only important that all the lines present in the observed
spectrum are also present in the template; it is also important that
the line depths in the template match the depths in the observed
spectrum. We establish the appropriate line depths in the template
spectrum in the following manner. As a starting point for this
iterative process we query the Vienna Atomic Line Database
\citep[VALD;][]{kupka1999} for line positions and line depths
appropriate for a star with the given effective temperature and
surface gravity (our inputs in this case are based on the SME
analysis; see Table~\ref{tab:koi244_overview}). We now adjust the line
depths so that a kernel convolved with the line list gives the best
fit (smallest $\chi^2$) to the high-S/N spectrum we had obtained
above. This kernel is created simultaneously with the optimization of
the line depths. At this stage the kernel is purely phenomenological,
and is not subject to any physical boundary conditions. It consists of
a number of free parameters. Each parameter represents one pixel along
the dispersion direction of the spectrograph, translated into velocity
for the wavelength region of interest ($\approx 1.3$~km\,s$^{-1}$ at
$390-480$~nm). The number of pixels is chosen such that the range of
velocities that is covered is about twice as large as the $v\sin
i_{\star}$ of the star. For HAT-P-2 we have $61$ free parameters and
for Kepler-25 we have $33$ free parameters. To increase the speed of
the computation each order is split into several sections, and for
each section the best fitting line depths and kernels are found
separately.

After this initial round we create an average kernel out of all
kernels from all sections in all orders. Here the kernels from each
section are weighted by the blaze function (for which we use the
B-star spectra as proxy) of that section, and the equivalent width of
the absorption lines in that section. We exclude regions at the short
and long wavelength end of each order, regions dominated by deep,
large lines, and regions for which clearly no good fit was achieved,
i.e.\ lines are missing in the template. About 10\% of the available
spectral range was excluded for one or another of these reasons.

This newly obtained average kernel is now used for all sections in a
second round, during which only line depths are adjusted. In the
following round only the section kernels are optimized. These last two
steps are repeated two more times to optimize the line depths. Finally
the template with the optimized line depths is saved for later use,
while the high-S/N spectrum as well as the average kernel are
discarded.

\begin{figure*}
  \begin{center}
\includegraphics[width=13.5cm]{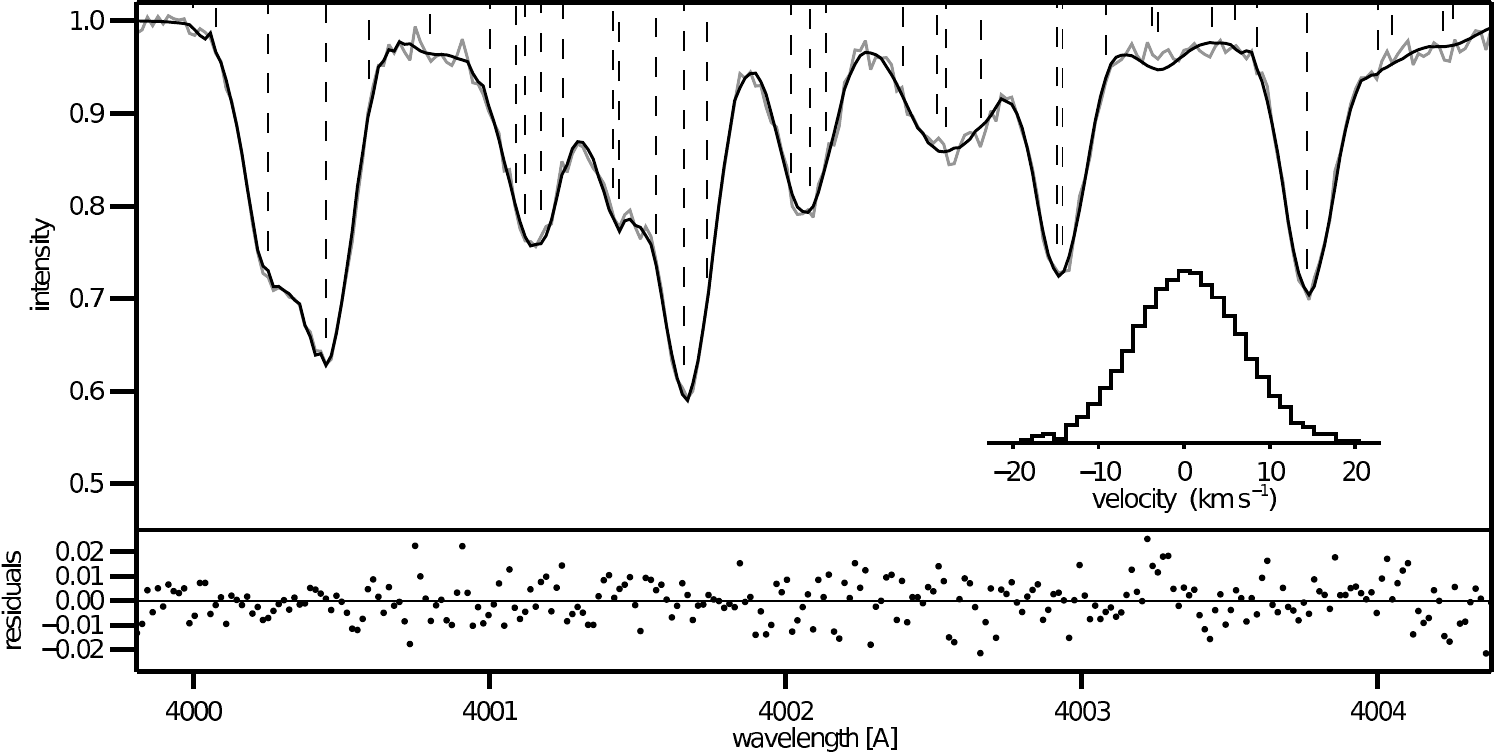}
\caption {\label{fig:koi244_spectrum} {\bf Small portion of the
    normalized spectrum of Kepler-25}. The thin gray line represents a
  small part of the third observed spectrum during the night of
  2012~May~31/June~1.  The dark line represents the convolution of the
  template (marked by the dashed lines) with the optimized kernel. The
  optimized kernel is shown in the lower right corner. The lower panel
  shows the difference between the observed spectrum and template
  convolved with the kernel. For each spectrum of Kepler-25, a total
  of about 900\,\AA\ were analyzed in the same fashion as shown in
  this 4\,\AA\ section.}
  \end{center}
\end{figure*}

\paragraph{Measuring high S/N ratio kernels.} Now we use the
depth-optimized template to obtain an average absorption line kernel
in each section of all observations.  Figure~\ref{fig:koi244_spectrum}
displays one section of one Kepler-25 observation during the transit
night. Also shown is the kernel derived from this section. Using the
same weighting scheme as used above we can now create an average
kernel for each observation.

\subsubsection{Analyzing high-S/N kernels}

The kernels obtained in the last section are then analyzed with the
same code which was used by \cite{albrecht2007}. In short we pixelate
the visible stellar surface and assign to each pixel a radial
velocity, based on contributions from stellar rotation,
macroturbulence, and the convective blueshift. At each phase of the
transit we integrate over the exposed portion of the stellar surface
to obtain the stellar absorption-line kernel. In this step we assume a
quadratic limb-darkening law. Finally the absorption line is convolved
with a Gaussian function representing both micro-turbulence in the
photosphere and the point-spread function (PSF) of HIRES.

\paragraph{Changes in the spectrograph PSF during the transit nights.}
Before the model absorption lines can be compared to the measured
kernels one last step has to be taken which is specific to these
observations. The spectra are obtained with a slit spectrograph and we
therefore have to take into account possible changes in the PSF
throughout the night. This is because small changes in telescope
guiding cause variability in the illumination of the slit, and
consequent changes in the PSF of the spectrograph.\footnote{In
  principle some information about the time-variable PSF is contained
  in the solutions provided by the RV-measuring code, which is based
  on a fit to the spectra in the iodine wavelength range from
  500-600~nm. However we did not thoroughly investigate such an
  approach as the PSF is also expected to vary with position on the
  spectrograph CCD.} Changes in illumination of the slit might shift
the PSF, sharpen or widen it, as well as introduce skewness and higher
frequency terms. As a measured spectrum is convolved with the PSF such
changes do effect the measured absorption lines directly.

We are interested in a time-varying signal: the distortion due to the
transiting planet. Therefore it is important to compensate for changes
in the absorption lines due to changes in the PSF. It is not crucial
to know the PSF itself. In the next paragraph we explain how we deal
with these potential shifts and stretches, and how we deal with
higher-order changes by interpolating between observations obtained
just before and after transit.

To compensate for changes in the absorption lines caused by changes in
the PSF we performed the following steps. We take the mean of the
first few spectra, and the mean of the last few spectra, during a
transit night. (If the PSF of the spectrograph would have been stable
and our correction for any RV changes would have been perfect, then
these should be identical, assuming no stellar activity). Next we
linearly interpolate in time between these to create an absorption
line appropriate for the time of each individual observation. These
lines do not contain a transit signal, but include slow monotonic
changes in the PSF. To isolate the transit signal we subtract the
measured kernels from these interpolated lines.

\begin{figure*}
  \begin{center}
         \includegraphics[width=6.5cm]{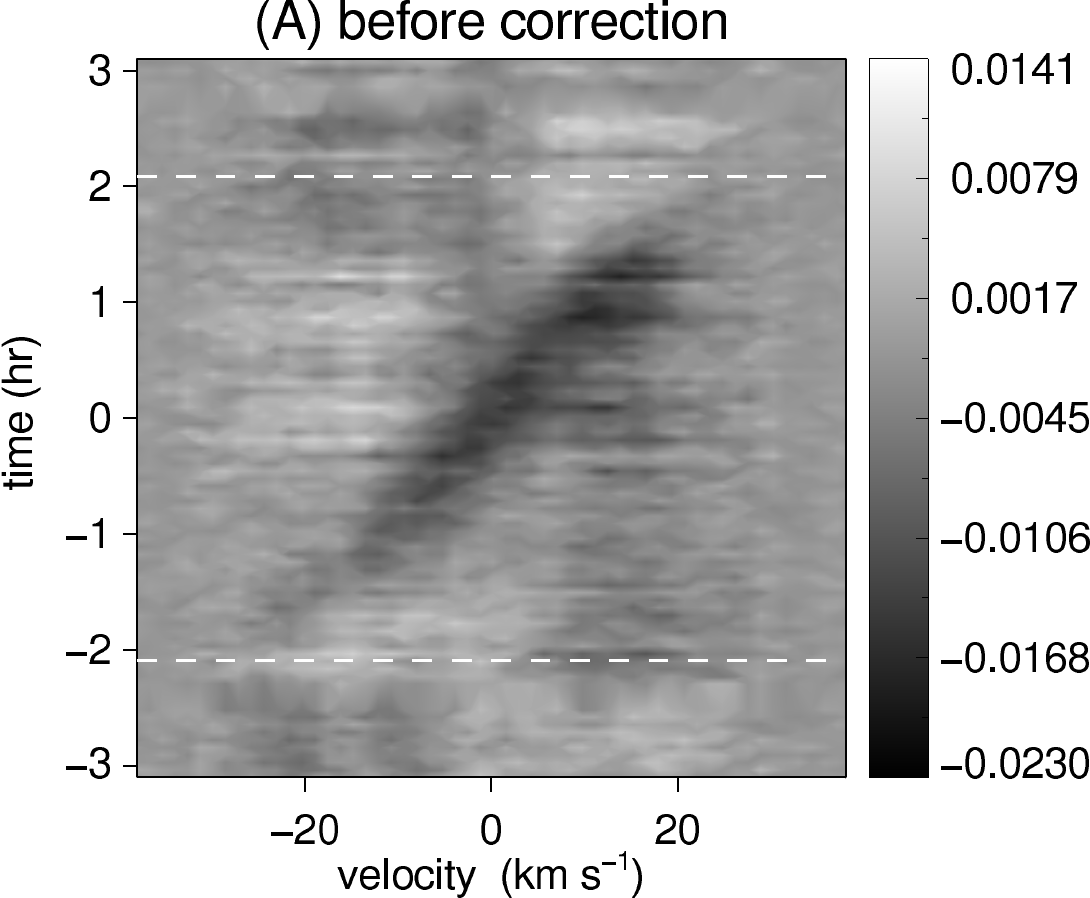}   
      \includegraphics[width=6.5cm]{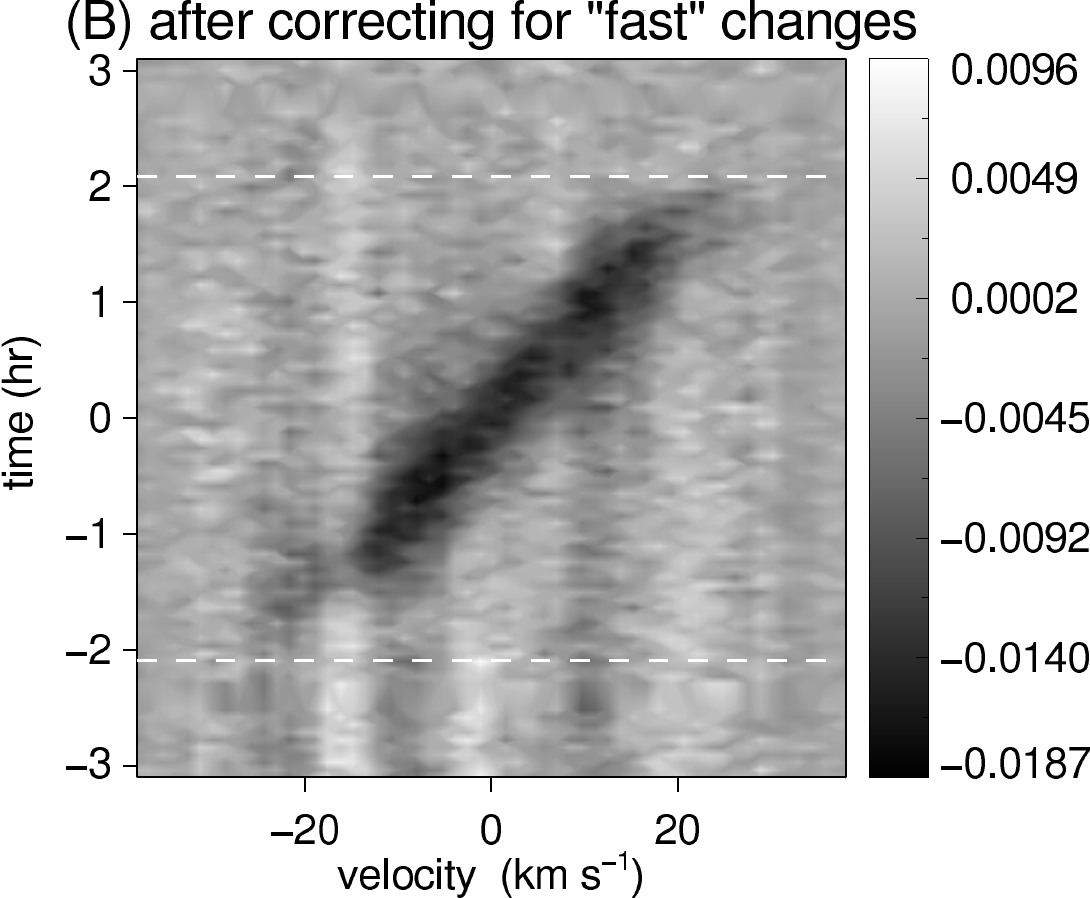}
    \includegraphics[width=6.5cm]{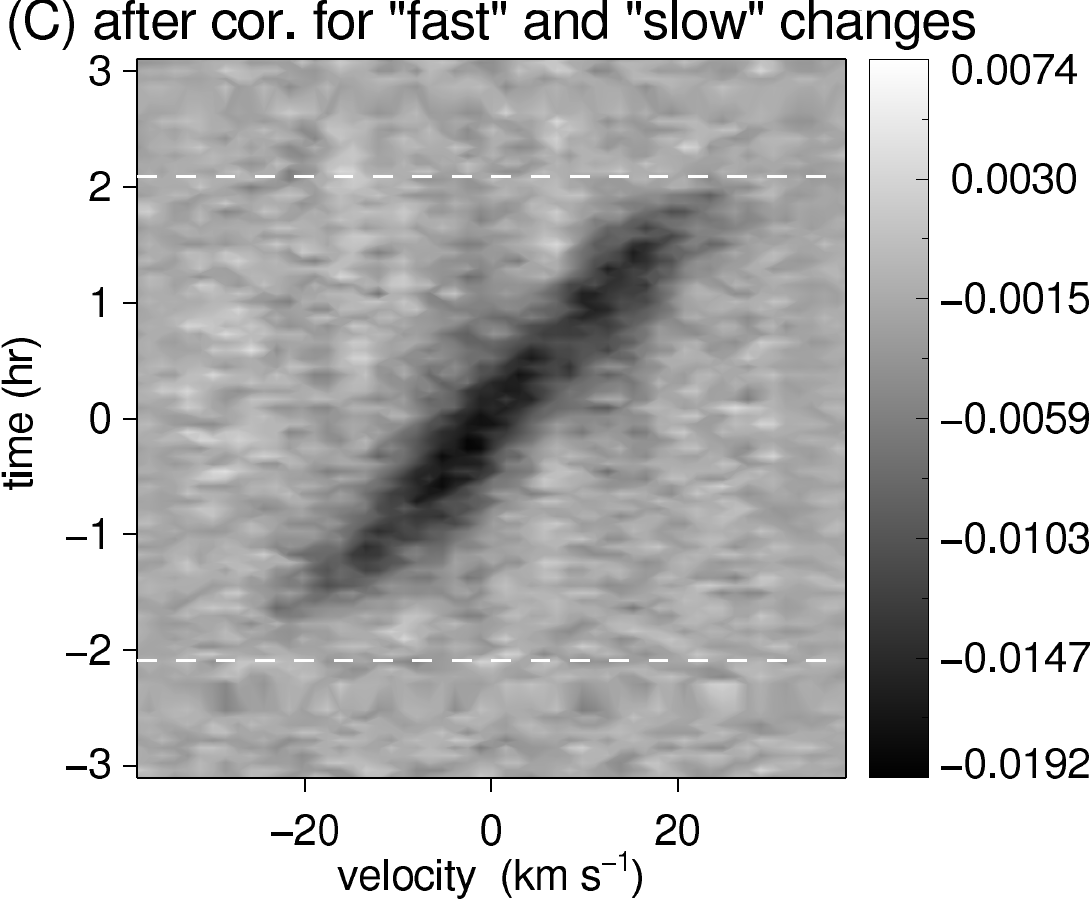}
      \includegraphics[width=6.5cm]{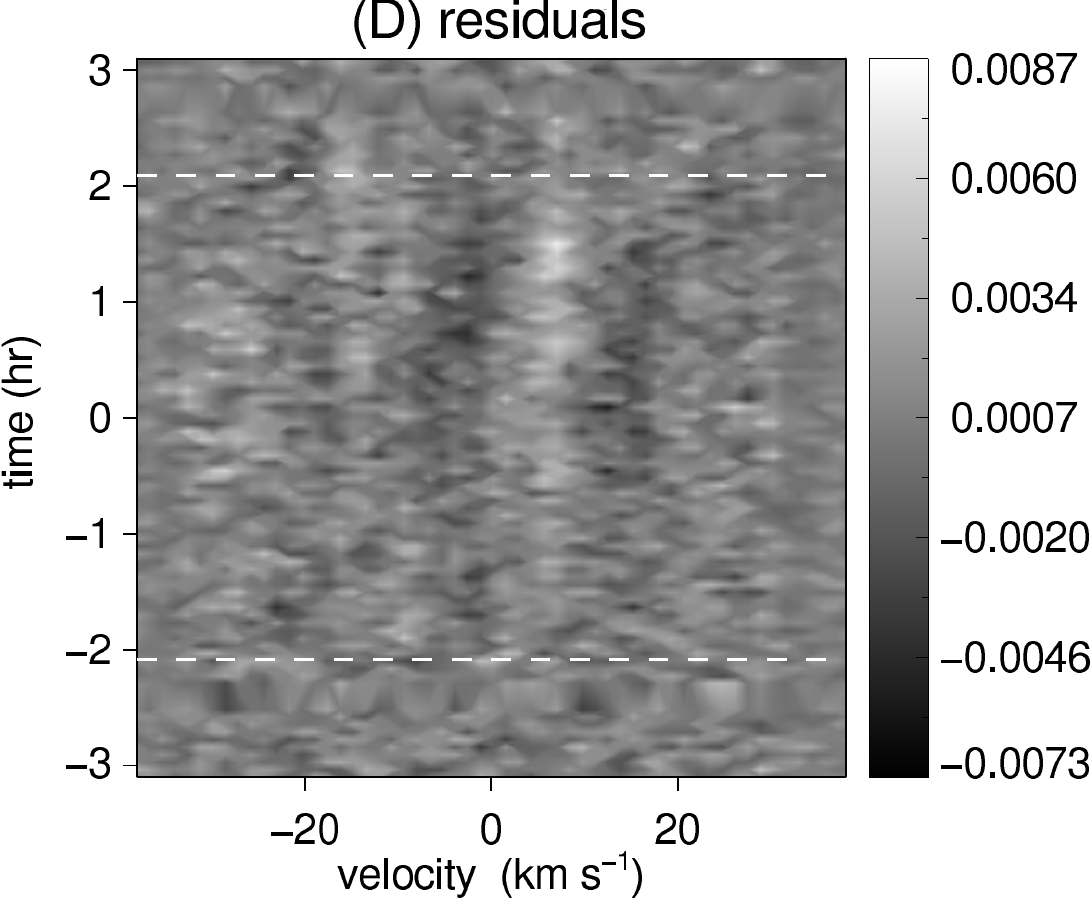}
      \caption {\label{fig:hd147506_transit} {\bf Doppler shadow of
          HAT-P-2b.}  During the planetary transit, part of the
        stellar photosphere is hidden from view, distorting the
        rotationally-broadened stellar absorption lines. The left
        upper panel ({\it A}) shows the time-variable planet shadow
        during a transit in the HAT-P-2 system. The dashed lines
        indicate times of first and last contact.  At the beginning of
        the transit, parts of the approaching stellar surface area are
        hidden from view, and therefore blueshifted light is
        hidden. At the end of the transit, redshifted light is hidden.
        In addition to the transit signal, some artifacts are visible.
        They originate mainly form hour-to-hour (``fast'') changes in
        the wavelength position of the kernels. A mismatch in
        wavelength between the out-of-transit kernel and the current
        kernel leads to a deficit on one side of the kernel (dark) and
        a positive residual on the other side (light color). Panel
        ({\it B}) shows the results after correcting for these fast
        changes, as described in the text. Here, for illustration
        purposes, we only used an average kernel based on the
        post-transit data, rather than the interpolation between
        pre-transit and post-transit kernels, as described in the
        text. There is a continuous buildup of difference between the
        observed and assumed kernels towards the beginning of the
        night (``slow'' changes). Panel ({\it C}) shows the results
        when the assumed kernel is based on interpolation in time
        between the pre-transit and post-transit kernels. Panel ({\it
          D}) shows the residuals after our best fitting model for the
        planetary transit is subtracted from the data shown in panel
        ({\it C}). There are some low-level patterns in the residuals,
        likely originating form changes in the PSF which are not
        modeled by our algorithm.  To facilitate comparison between
        the different panels, the kernels form each system have been
        normalized to a height of unity (see
        Figure~\ref{fig:koi244_shape} and
        Figure~\ref{fig:average_line}).  The grayscale bar next to each
        panel indicates the signal strength on the same scale.
        It is interesting to note that the depth of the
        HAT-P-2b velocity signal is more than twice as deep as the
        photometric transit signal. For rapidly rotating stars the
        depth of the deformation is not proportional to the square of
        the radii ratio, but is better approximated by the ratio
        itself.}
  \end{center}
\end{figure*}
\begin{figure*}
  \begin{center}
    \includegraphics[width=6.5cm]{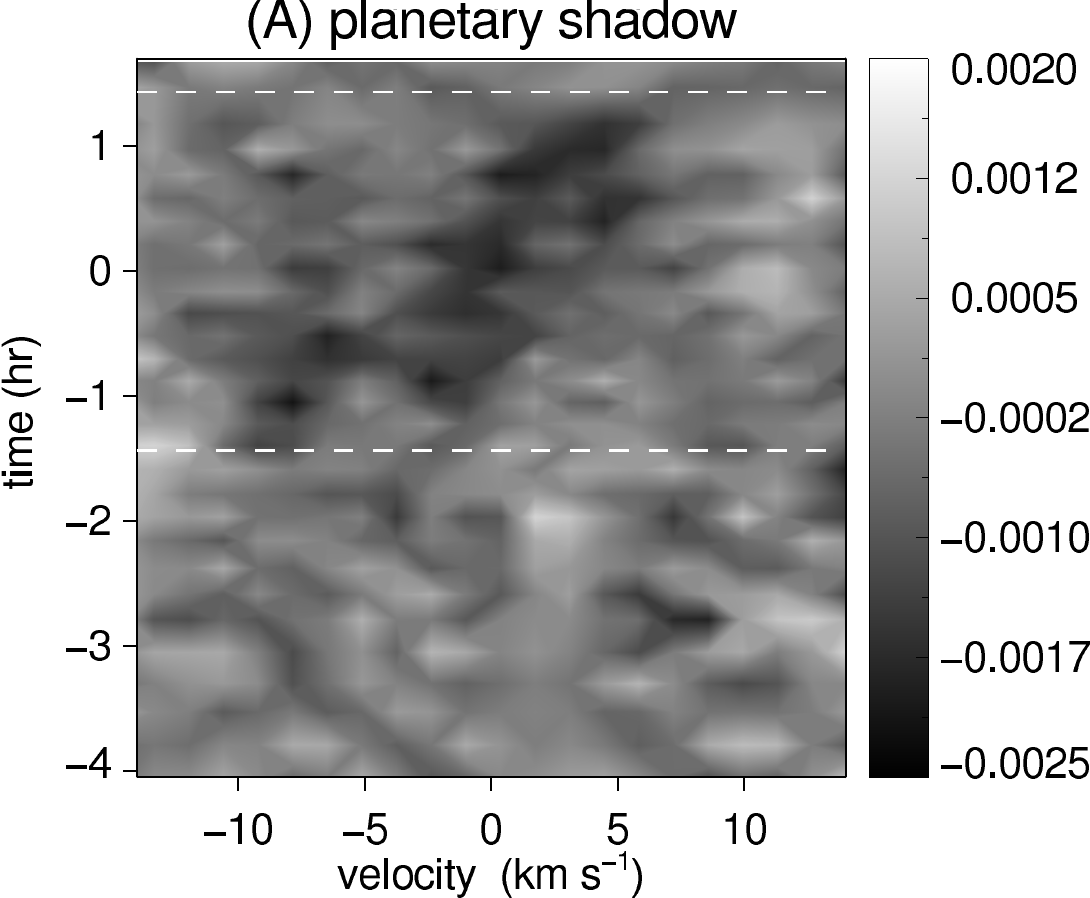}
    \includegraphics[width=6.5cm]{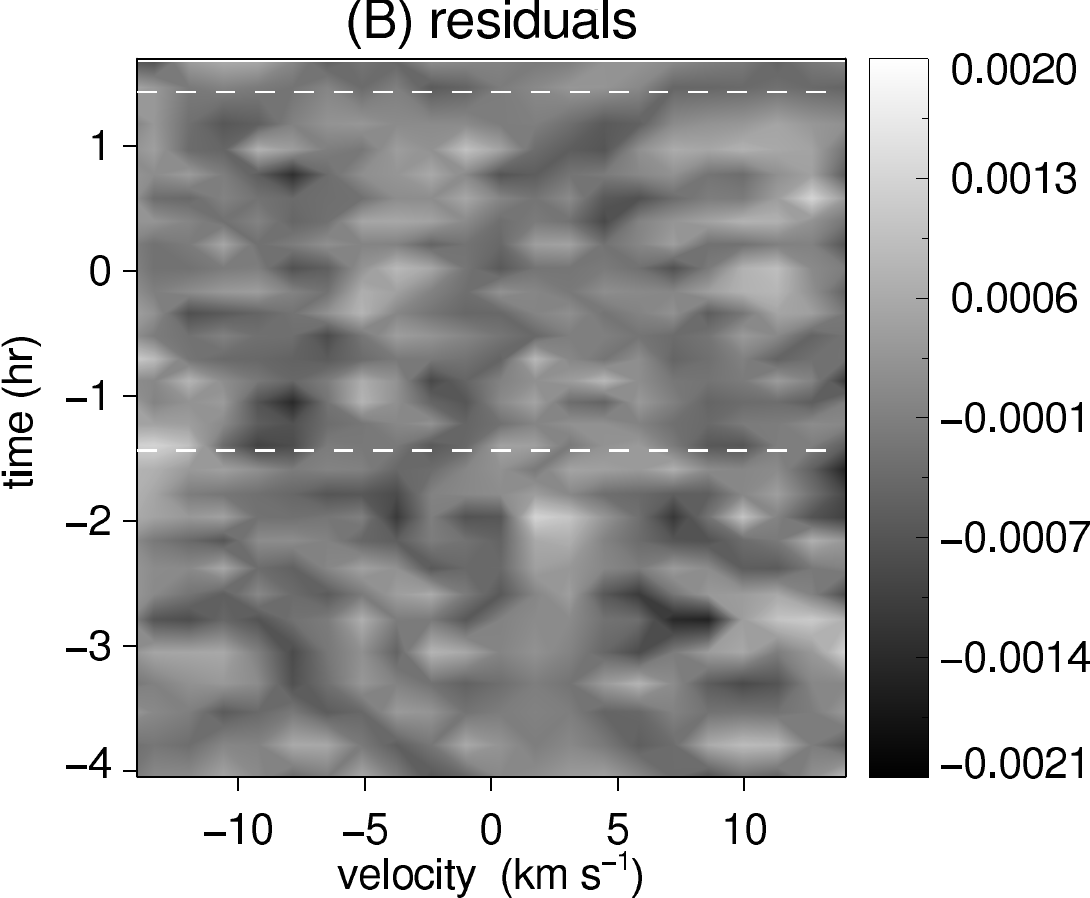}
    \caption {\label{fig:koi244_shape} {\bf Doppler shadow of Kepler-25c.}
      The same as panels (c) and (d) in
      Figure~\ref{fig:hd147506_transit}, but this time for the transit
      on 2012~May~31/June~1 of Kepler-25c. In the left panel (A)
      one can see the signature of a distortion traveling from
      negative RV towards positive RV throughout the transit: the signature
      of a prograde orbit. Modeling the distortion gives
      $\lambda=-0.5 \pm 5.7^{\circ}$, indicating good alignment between the
      projections of the stellar and orbital spins. Panel (B) shows
      the residuals after subtraction of our best-fitting model.}
  \end{center}
\end{figure*}

To correct for low-order fast changes in the PSF we allow the kernel
of each observation to shift in velocity space and we further allow
for a scaling in the velocity scale of the measured absorption
line. This leads to two free parameters for each observation, which
are evaluated each time the observations are compared to a specific
model. The drifts are less then $200$~m\,s$^{-1}$ and the scaling in
velocity space is always less than $0.5$\,\%. See
Figure~\ref{fig:hd147506_transit} for an illustration of how these
different corrections influence the signal.

With this scheme we can compensate for a constant unknown PSF, slow
changes in the PSF, as well as fast low-order changes in the PSF. We
do not attempt to correct for fast high-order changes in the PSF as
these would likely be correlated with the planet transit signal, which
is itself a higher-order change in the stellar absorption lines.

\begin{figure}
  \begin{center}
    \includegraphics[width=8cm]{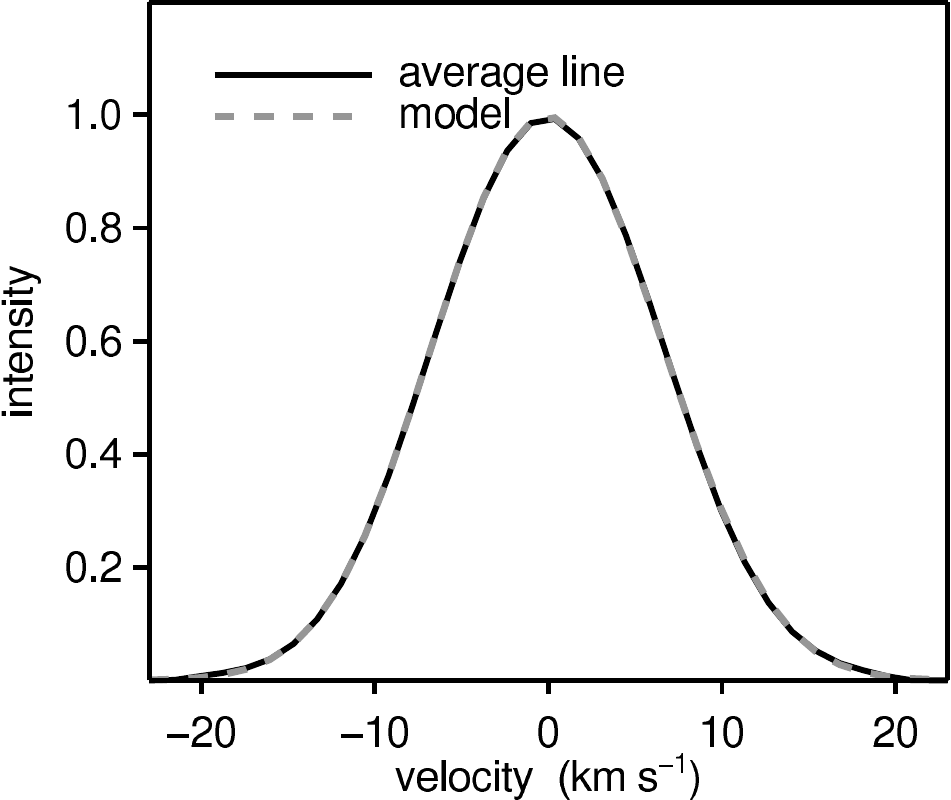}
    \caption {\label{fig:average_line} {\bf Comparison of overall line shape to
        model.} The solid line shows the
      average line shape of Kepler-25 as measured with the  first
      three observation during the transit May/June 31/1, 2012. The
      gray dashed line shows our best-fitting model of this line. The model includes
      quadratic limb-darkening, stellar rotation, micro and macroturbulence, 
      the convective blueshift, and a PSF which is modeled as a simple Gaussian. 
      Apart from the $\sigma_{\rm PSF}$ parameter, all other
      parameters are also simultaneously used to find the best-fitting
      model for the planet shadow (Figure~\ref{fig:koi244_shape}). Our results for HAT-P-2 are not shown
      here but are of similar quality.}
  \end{center}
\end{figure}

In addition to analyzing the changes in the absorption lines, we also
use the observations taken before transit and compare them to the
model line.  This gives additional constraints on \vsini, $\zeta$, and
limb darkening. Here the PSF is modeled as a simple Gaussian with only
one free parameter, the width ($\sigma_{\rm PSF}$). No accurate
modeling of the PSF is required as we do not attempt to identify a
small transit signal, but rather simply to measure the line width.
See Figure~\ref{fig:average_line} for a comparison between a measured
absorption line and our model. The evaluation of the planet shadow and
the out-of-transit line is done simultaneously.

\begin{figure*}
  \begin{center}
    \includegraphics[width=12.5cm]{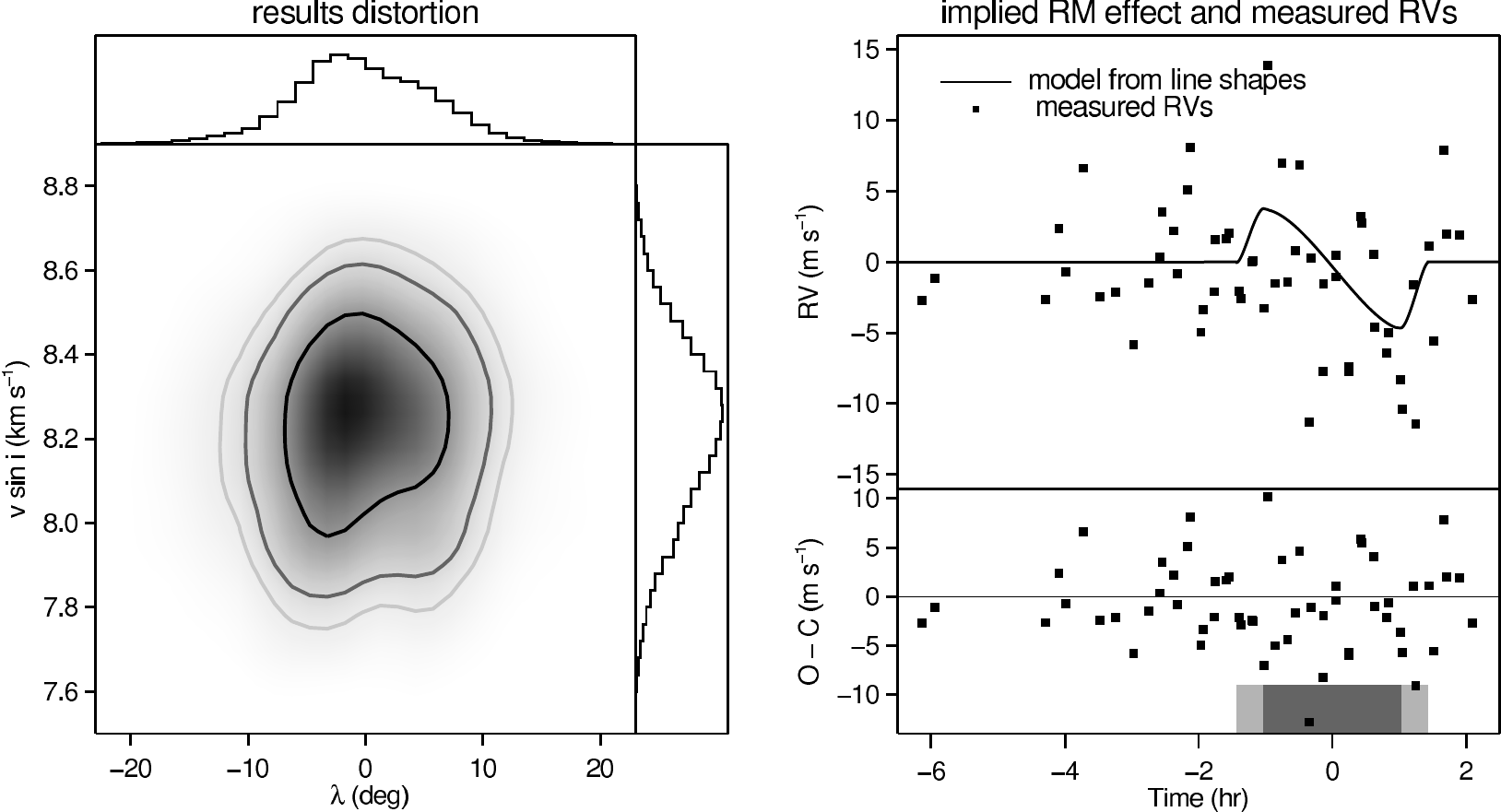}
    \caption {\label{fig:koi244_shape_results} {\bf Kepler-25 results
        form the analysis of the absorption lines and comparison to
        the RVs.} The left panels show our results from the analysis
      of the distortion in the stellar absorption lines. We measure
      $\lambda=-0.5\pm7^{\circ}$ and $v \sin
      i_{\star}=8.2\pm0.2$\,km\,s$^{-1}$.  To allow a comparison to
      the RV data, the solid line in the right panel shows the
      anomalous RV signal that is {\it implied} by the best-fitting
      model of the line distortions. It is seen here to be compatible
      with the RV data even though the RV data was not used directly
      to constrain this model.  (For this visual comparison, the
      out-of-transit RV trends were subtracted from the measured RVs.)
      The lower panel shows the difference between the measurements
      and the model, illustrating the good agreement between the
      line-distortion method and the anomalous-RV method for
      characterizing the planetary transit.}
  \end{center}
\end{figure*}

\paragraph{HAT-P-2} As our technique had to this point only been used
for binary star systems, we applied it to the HAT-P-2 system before
using it on the more challenging Kepler-25 dataset. The spectra of the
HAT-P-2 system were obtained and analyzed by \cite{winn2007} and the
RVs were reanalyzed by \cite{albrecht2012b} who found a low projected
obliquity ($\lambda=9\pm10^{\circ}$; the formal result was
$\lambda=9\pm5^{\circ}$, but due to residual structure in the RVs
after substruction of our best fitting model we estimated the true
uncertainty to be higher). Here we use the spectral region from $443$
to $455$~nm and the same photometric priors as used by
\cite{albrecht2012b}. We find $\lambda=7.6\pm0.5^{\circ}$. This is
compatible with the RV-based results and, formally, implies a small
misalignment in this system. However, given the patterned residuals
visible in panel D of Figure~\ref{fig:hd147506_transit} the true
uncertainty is probably larger. To investigate this very high-S/N
dataset further, the fidelity of our spectral model would need to be
increased, and it would also be preferable to repeat the measurement
with a different spectrograph. If the misalignment is confirmed than
this would make HAT-P-2 an important system to study tidal
alignment. Here, with the good agreement between the results of the RV
and shape methods, we gained confidence that our algorithm to extract
projected obliquities directly from modeling stellar absorption lines
also works for the case of planetary transits observed with a slit
spectrograph.

\subsubsection{Kepler-25: changes in stellar absorption lines}
\label{sec:koi244_shape}

Using the scheme described above we then analyzed the Kepler-25
spectra obtained during the two transit nights. We used the spectral
region from $398$ to $479$~nm, blueward of the iodine lines. We did
this separately for each transit night. We first focused on the
transit night 2012~May~31/June~1. We used the same composite
photometric light curve as used in our analysis of the RVs, and left
$P$ fixed at the value determined from the photometry. With this
method of analysis there is no need to determine $K_{\star}$ and
$\gamma$ for the transit nights, as had been necessary for our
analysis in Section~\ref{sec:koi244_rvs}. This reduced the number of
free parameters by two; on the other hand there was a new parameter
$\sigma_{\rm PSF}$ as described above. We did not impose any prior
constraints on this parameter. To estimate the parameters and their
uncertainty intervals we again used an MCMC algorithm.

\begin{figure}
  \begin{center}
    \includegraphics[width=8.0cm]{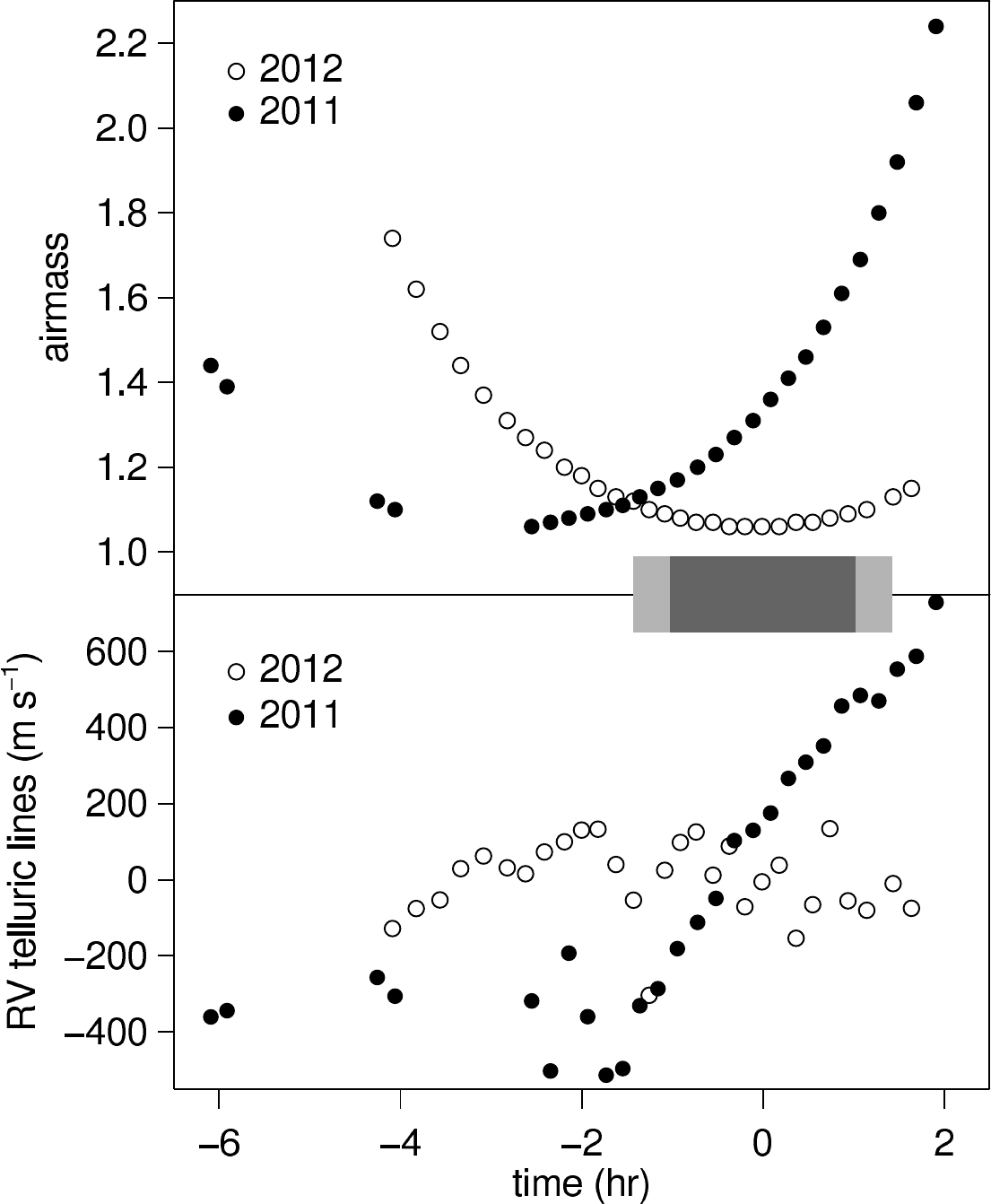}
    \caption {\label{fig:koi244_environment} {\bf Changes in airmass
        during the Kepler-25 observations.} The upper panel shows the
      changes in air-mass for the observations in 2011 (filled symbols)
      and 2012 (open symbols). The transit interval is
      indicated by the gray bars. {\it Lower panel:} The measured RVs
      for telluric lines on the red CCD of HIRES. }
  \end{center}
\end{figure}

\begin{figure*}
  \begin{center}
    \includegraphics[width=18.cm]{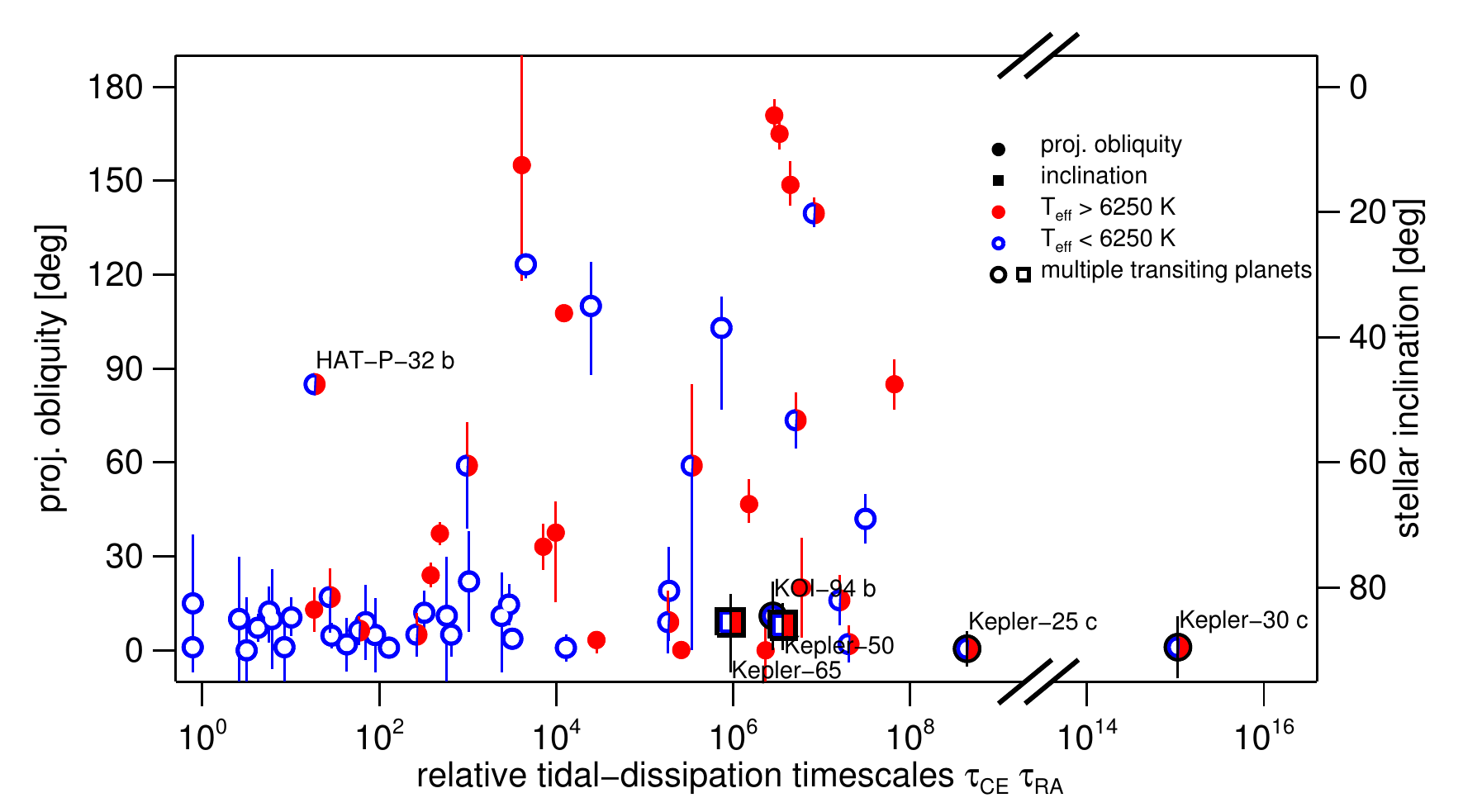}
    \caption {\label{fig:tau1} {\bf Projected obliquity (either
        $\lambda$ or $i_\star$) as a function of the relative tidal-alignment timescale,
        for hot-Jupiter and multiple-planet systems.}
      The systems are plotted as a function of a simple metric for the
      expected timescale for tidal dissipation within the star
      (see \citealt{albrecht2012b} eq.~2--3). Stars which have
      temperatures higher then $6250$~K are shown with red filled
      symbols. Blue open symbols show stars with temperatures lower
      then $6250$~K. Stars which measured effective temperature
      include $6250$~K in their 1-$\sigma$ interval are shown by split
      symbols. Systems for which $\lambda$ was measured are indicated
      by a circle and refer to the left-hand axis.
      Measurements of $i_{\star}$ are indicated by a square and refer
      to the right-hand axis. Systems which harbor multiple planets
      are given dark black borders. The systems with short tidal
      timescales are seen to be well-aligned. 
      All of the multiple-planet systems are well-aligned despite
      having weak tidal dissipation.}
  \end{center}
\end{figure*}

Figures~\ref{fig:koi244_shape} and \ref{fig:average_line} show the
comparison between the data and the best-fitting models. Our results
for \vsini\ and $\lambda$ are shown in the left panel of
Figure~\ref{fig:koi244_shape_results} and printed in column 3 of
Table~\ref{tab:koi244_results}. These are consistent with the results
found using the RVs of both data sets (Section~\ref{sec:koi244_rvs},
Table~\ref{tab:koi244_results} column 2). In particular the results
for $\lambda$ are consistent with each other. Analyzing RVs obtained
with the iodine technique we obtain $\lambda=5\pm8^{\circ}$. Analyzing
the change in the absorption lines in the blue part of the spectrum we
obtain $\lambda=-0.5\pm5.7^{\circ}$. These results are independent
from each other as different wavelength regions of the obtained
spectra have been used (although the supporting photometric data was
the same in both cases). In addition, the two independent measurements
of \vsini are consistent with each other. To illustrate the
consistency of the results we show in the right panel of
Figure~\ref{fig:koi244_shape_results} the expected RV signal from our
best solution to the distortion of the absorption lines in the blue
spectra. We further show the RVs measured during the two transits in
the red part of the spectra, and which are displayed in
Figure~\ref{fig:koi244_rv_results}. We also show the difference
between the measured RVs and the implied RV changes from the
distortion. These signals are simply plotted on the same axes; they
were not adjusted to match each other. The agreement of these two
different approaches lends additional confidence to the conclusion
that the projected stellar obliquity is low in Kepler-25. Why do we
obtain a smaller uncertainty interval for $\lambda$ by measuring the
deformations of the lines, rather than measuring RVs? We believe there
are two reasons. When the line width is dominated by rotation, the
spectroscopic transit depth is deeper than the broad-band transit,
because in the spectroscopic transit only the portion of the star with
the appropriate radial velocity is contributing to the contrast
(Figure~\ref{fig:koi244_shape}). Furthermore the $\lambda$ parameter
is not strongly correlated with the other parameters that alter the
position of the spectral lines, unlike the strong correlations that
are observed when fitting RV data.

However, for the first transit night (2011~July 18/19) there was no
secure detection of the Doppler shadow. What might have prevented a
detection in this case? The main difference between the two datasets
is a difference in the range of airmasses through which the
observations were made (Figure~\ref{fig:koi244_environment}, upper
panel). During the 2011 transit the airmass increased from $1.1$ to
$1.9$. In contrast, in 2012 the transit was observed at low zenith
angles, always below an airmass of 1.1. The large variation in airmass
during the 2011 observations strongly increases the difficulty of
modeling the observed spectra. This is because a change in the angle
under which an object is observed can lead to changes in the slit
illumination which in turn changes the effective PSF. This can most
readily be seen in the lower panel of
Figure~\ref{fig:koi244_environment} where the apparent shift in radial
velocity of telluric lines is plotted against time. The apparent
changes are about $\approx1$~km\,s$^{-1}$ for the night where we
detect no signal and only $\approx150$~m\,s$^{-1}$ where the changes
in airmass are low and where the transit signal was detected. Telluric
lines are intrinsically stable to at least a few tens of m\,s$^{-1}$
\citep[e.g.][]{gray2006,figueira2010}.

The apparent changes in radial velocity of the telluric lines are
measurement artifacts, caused by changes in the PSF, which also apply
to stellar lines. We suspect that on top of the RV shifts higher order
changes occur, which our algorithm cannot correct with sufficiently
accuracy to allow for the detection of the small distortion induced by
the planetary transit. During the observations of our test system
HAT-P-2 the air-mass also changed significantly, from 1.2 to 2, and
the observed RVs of telluric lines changed by $\approx
1$~km\,s$^{-1}$. The residuals of $0.8$~\% in our test system
(Figure~\ref{fig:hd147506_transit}, panel D) are larger than the
expected distortion of $0.2$~\% in the absorption lines in Kepler-25
(Figure~\ref{fig:koi244_shape}, panel A). The signal of HAT-P-2b was
nevertheless detected because of its relatively large amplitude; but
the nondetection of a transit signal during the 2011 observations of
Kepler-25 is not surprising in this context.

\section{Comparison with Hot-Jupiter systems}
\label{sec:comparison}

In this section we will analyze our results for KOI-94 and Kepler-25
together with the results for three additional multi-transiting
systems to try and clarify the interpretation of the high obliquities
seen in hot-Jupiter systems. The other three systems are Kepler-30,
50, and 65. Using the occurrence of star spots, \cite{sanchis2012}
measured $\lambda$ to be $1\pm10^{\circ}$ in
Kepler-30. \cite{chaplin2013} measured the inclination of the stellar
rotation axis towards the observer ($i_{\star}$) for two additional
systems, via asteroseismology.  For Kepler-50 they found
$i_{\star}=82^{+8}_{-7}$$^{\circ}$ and for Kepler-65 they found
$i_{\star}=81^{+9}_{-16}$$^{\circ}$. An estimate of the stellar
inclination in Kepler-50 had also been obtained earlier by
\cite{hirano2012b}, using a combination of estimates for the rotation
period, stellar radius, and $v\sin i_\star$.  Their result was less
constraining than, but compatible with, the result of
\cite{chaplin2013}.
 
\paragraph{High obliquities: planet migration or star-disk evolution?}
For HJ systems, evidence has accumulated that the stellar obliquities
varied over a very large range when the gas giants arrived near their
host stars \citep{winn2010,schlaufman2010,albrecht2012b}. This has been taken as
evidence that the orbital plane of the planet has changed, presumably
via the same mechanism which also changed its orbital
distance. However, as mentioned in the introduction, there are other
mechanisms which might create a misalignment between the stellar
orientation and the planetary orbit. In multiple-transiting planet
systems there is reason to think that the orbits still trace the plane
of the disk out of which they have formed. Therefore measuring the
obliquities in these systems lets us learn about the degree of
alignment between protoplanetary disks and stellar spin axes.

If we find that the distribution of obliquities for multiplanet
systems is closer to alignment, than the distribution of obliquities
for HJ systems, then this would indicate that the large obliquities in
HJ systems are caused by the evolution of the planets' orbits. If on
the other hand we find that the distribution in obliquities for
multiplanet systems is similar to the distribution of obliquities for
HJs then the measured obliquities are not necessarily related to
planet migration.

We note that the host stars in both groups, close in gas giants and
multiple planet systems, are on the main-sequence and cover the
spectral classes from F to K. The only readily apparent difference
between these systems is that for one group, several planets are found
on compact coplanar orbits, within in the other group there are
solitary transiting gas giants.

\paragraph{The influence of tides.} Before we can compare the two
distributions of obliquities we first need to know if tides have
dampened the obliquities. It would be advantageous to only include
systems which have not undergone any significant tidal influence,
instead of attempting to model the influence of tides on the
obliquity.

To check which systems might be influenced by tides, we calculate the
same two alignment timescales presented in \cite{albrecht2012b} for
the multiple planet systems. Either approach to calculating the
timescale leads to the conclusion that tides probably had little
or no influence on the stellar obliquities in the five multiplanet
systems. In Figure~\ref{fig:tau1} we show the results for the
timescale which is calibrated based on binary-star data
\citep{albrecht2012b}. In the same Figure we also show all the
hot-Jupiter systems from \cite{albrecht2012b}, after including some
newly published measurements. We included the new measurements for
WASP-32 and WASP-38 from \cite{brown2012}, HAT-P-17 \cite{fulton2013},
updated the value for CoRoT-11 from \cite{gandolfi2012}, and updated
the value for WASP-19 from \cite{tregloan-reed2012}.\footnote{Recently
  \cite{hebrard2013} measured $\lambda$ in the
  WASP-52 system. We do not include this result in the current
  study. The reasons are similar to the reasons for which we excluded
  some systems in the \cite{albrecht2012c} study.} As mentioned we
also added the $\lambda$ and $i_{\star}$ measurements for the multiple
transiting planet systems. The left-hand $y$-axis indicates $\lambda$
and the right-hand $y$-axis indicates $i_{\star}$. Indication of good
alignment is a low value in $\lambda$ or a large value of
$i_{\star}$. The long tidal realignment timescales in the studied
multiple planet systems is due to the long orbital periods and the
small masses of the planets.

According to Figure~\ref{fig:tau1} it does seem unlikely that any of
the multiple planet system was influenced by tides. But which hot
Jupiter systems should be included in the comparison? As we do not
have a clear cut criterion we will use three samples: (1) All hot
Jupiters; (2) all hot Jupiters with $\tau$ equal or larger to the
$\tau$ of the first clearly misaligned system ($\tau>10^{2.7}$); and
(3) only hot Jupiters which have timescales $\tau$ equal to or larger
than the shortest $\tau$ found amongst the multiple-planet systems
($\tau=10^{5.8}$).

\begin{figure}
  \begin{center}
    \includegraphics[width=8cm]{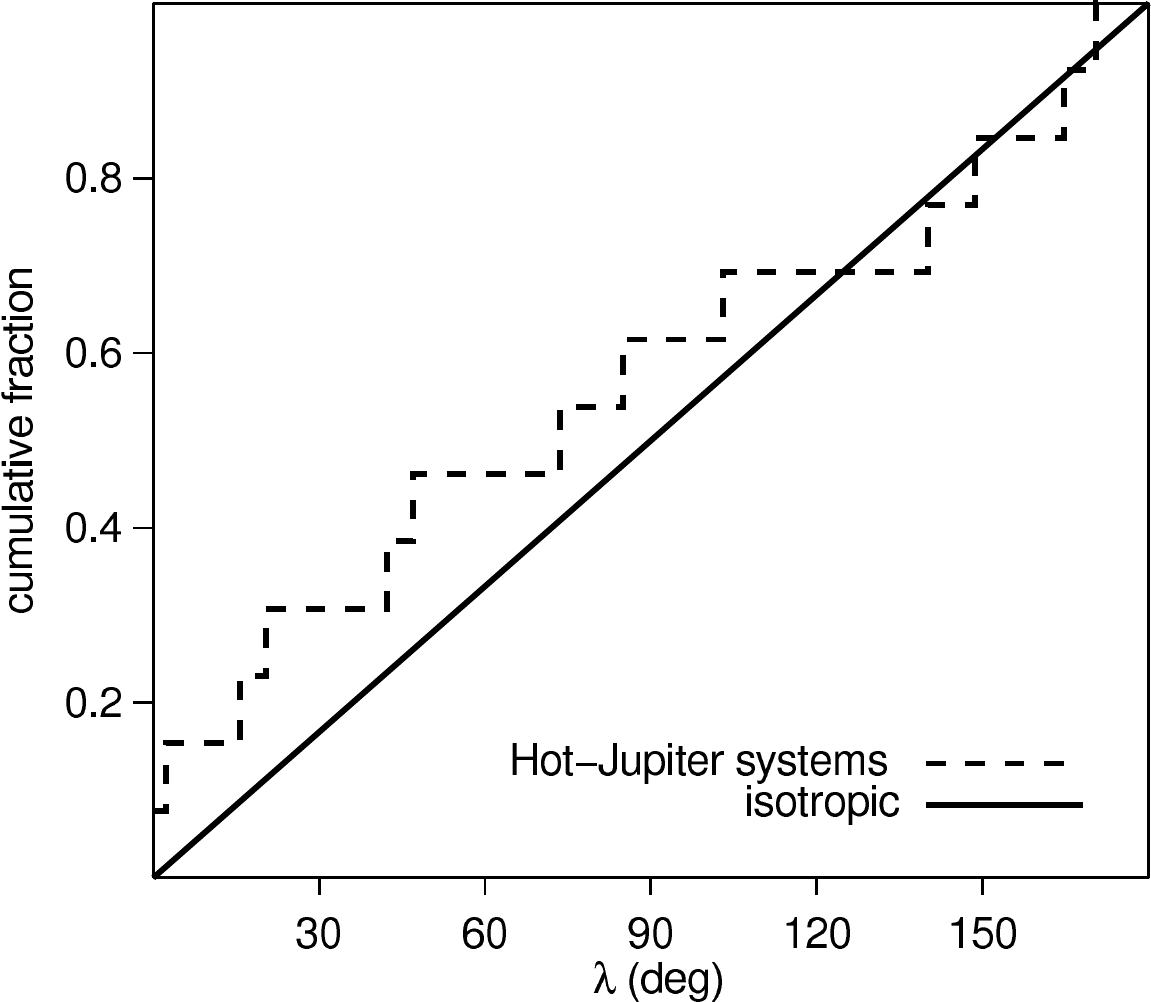}
    \caption {\label{fig:k_s} {\bf Kolmogorov-Smirnov test on the
        isotropic distribution of $\lambda$ for single, close in, gas
        giant planets.} The solid line shows the
      cumulative fraction of $|\lambda|$  for an isotropic
      distribution of obliquities on a sphere,  all $\lambda$ are equally likely {\it a priori}. The dashed line
      shows the cumulative distribution for measurements of $\lambda$
      in HJ systems. We only
      included systems with $\tau> 10^{5.8}$ (See
      Figure~\ref{fig:tau1}), to avoid systems which are
      strongly influenced by tides. According to the
      Kolmogorov-Smirnov test there is a $61$\% chance that
      the projected obliquities of these systems are drawn from an
      isotropic distribution.}
  \end{center}
\end{figure}

\paragraph{Comparing the distributions.} In the regime of weak tides,
the hot Jupiter results appear to be nearly random. Therefore we first
ask: could either population be drawn from an isotropic distribution
on a sphere?  For hot Jupiters we have only measurements of $\lambda$,
the projected obliquity. We can therefore compare these measurements
to a distribution in $\lambda$ for the isotropic case using a
Kolmogorov-Smirnov (K-S) test \cite[e.g][]{press1992}. For case (1)
the K-S test suggests that there is negligible probability that the
$\lambda$ measurements of all hot-Jupiter systems are drawn from an
isotropic distribution. For case (2) there is still only a $0.04$~\%
probability that the results are drawn from an isotropic
distribution. However for case (3) we find there is a
$61$~\% chance that this distribution is consistent with an isotropic
distribution in $\lambda$. Figure~\ref{fig:k_s} shows the cumulative
distribution in $\lambda$ for these HJs and the expected distribution
in $\lambda$ for an isotropic distribution.

For the multiple planet systems we have two distinct measures of
obliquity, $\lambda$ and $i_{\star}$, which cannot be translated into
each other (at least not without already assuming a distribution, see
\citealt{fabrycky2009}).  Therefore we use a Monte Carlo approach
instead of a K-S test.  We create a distribution of obliquities which
has a uniform distribution in $\lambda$ (Kepler-30, KOI-94, Kepler-25)
and in $\cos i_{\star}$ (Kepler-50, Kepler-65) to simulate a isotropic
distribution in the obliquities. Form these we draw five
``measurements'' which we compare to the three measurements of
$\lambda$ and two measurements of $i_{\star}$. The uncertainties in
the actual measurements are included as Gaussian random numbers, every
time a comparison is made. In particular for the comparison in
$i_{\star}$ we use half-Gaussians with peaks at $90^{\circ}$ and
standard deviations derived from the inclination measurement plus the
measurement uncertainty. We repeat this experiment $5\times10^{7}$
times. Only in $0.0003$\% of these experiments do we draw sufficiently low
values of $\lambda$, and sufficiently high values of $i_{\star}$,
to be compatible with the measured $\lambda$s and $i_{\star}$. It seems
unlikely that obliquities in multiple planet systems are drawn from an
isotropic distribution. A narrow distribution centered near zero
obliquity is more appropriate. We will defer an analysis of which is
the exact distribution until more obliquity measurements in
multiplanet systems are available. See \cite{fabrycky2009} for
possible approaches on how a comparison can be made. Such an analysis
would be interesting as it might shed light on the origin of the small
($6^{\circ}$) obliquity of the Sun.

Now combining that 1) multiplanet systems have a different obliquity
distribution than systems with single, close-in gas giants, 2) planets
in multiple planet systems presumably trace with their orbits the
plane of the circumstellar disk out of which they formed, and 3) we
are not able to detect any other significant difference between stars
which have close in giant planets and stars which hosts multiple
planets, we conclude that the misalignments between stellar rotation
and planetary orbits are due to changes of the inclinations of the
orbital planes. Our results disfavor theories which aim to explain large obliquities
due to a change in the angle between protoplanetary disk and the star
\cite[e.g.][]{bate2010,thies2011,batygin2012} or changes in the
internal structure of the star \citep{rogers2012}.

Of course it must be acknowledged that a sample of 5 systems is not
sufficient for a firm conclusion. The systems studied here only cover
a small parameter space, for example a limited range in stellar mass.
In other systems, mechanisms for tilting stars may be more
important. One clue that this is indeed the case is the finding that
both stars in the DI\,Herculis system are strongly inclined with
respect to their mutual orbit \citep{albrecht2009}. Here however, we
have found that the evidence to date supports the conclusion that the
high obliquities of hot-Jupiter systems are due to evolution of the
planets' orbits.

\paragraph{Relation to planet migration theories.} 
If we assume that the smaller mass planets in multiple-planet systems
migrated inwards then it seems that we have (at least) two types of
processes which may be of importance in planet migration. One type of
process changes the obliquity while the other does not. We could
identify disk migration with the latter, and dynamical interactions
with the former. We note, though, that it is not necessarily true that
the compact multiplanet systems have experienced inward migration;
see, for example, \citealt{chiang2012}.

A number of mechanisms have been proposed for changing the orbital
inclination of a planet. Two processes which have attracted particular
attention are planet-planet scattering
\citep[e.g.][]{rasio1996,chatterjee2008,matsumura2010,nagasawa2011}
and Kozai Cycles with Tidal Friction (KCTF, \citealt{eggleton2001}).
Kozai cycles can be induced by the influence of a distant stellar
companion \cite[e.g.][]{wu2003,fabrycky2007,naoz2012} or by a distant
planet \cite[e.g.][]{naoz2011}. To confirm Kozai migration via a
stellar companion, searches for stellar companions to hot-Jupiter
hosts will be helpful \citep[e.g.][]{narita2012}. Another way to test
theories of hot Jupiters involving tidal circularization from a highly
eccentric orbit is to search for their putative high-eccentricity
progenitors.  \cite{socrates2012} predicted that there should be a
stream of gas giant on very eccentric orbits (eccentricity $<0.9$) if
hot Jupiters were transported directly inwards from beyond the snow
line. \cite{dawson2012} searched the {\it Kepler} database for such
objects, did not find any, and placed an upper bound on such a
population.  This indicates that if KCTF is an important migration
path, then likely the starting point for KCTF is closer than the snow
line.  This suggests that gas giants migrate via a combination of
processes. For example initial scattering or disk migration followed
by Kozai cycles and finally tidal circularization.

\section{Summary}
\label{sec:summary}

In the multiple-planet systems KOI-94 and Kepler-25 we measured good
alignment between the stellar rotation axes and the orbital plane of
the transiting planets. For both systems we used the radial velocity
anomaly (Rossiter-McLaughlin effect) during planetary transits to
determine the degree of alignment. For KOI-94 our result is consistent
with an independent study by \cite{hirano2012}. As the
Rossiter-McLaughlin effect in the Kepler-25 system has only a small
amplitude, we further measured the distortion of the stellar
absorption lines directly in another part of the obtained stellar
spectra. We found consistent results with both methods.

Combining our results with measurements in three other multiple planet
systems (Kepler-30, Kepler-50, Kepler-65) we can now compare the
obliquity distributions of multiple planet systems to the obliquities
measured in Hot-Jupiter systems. We find that there is only a
$0.0003$\,\% chance that multiple planet systems have a isotropic
obliquity distribution. This is in contrast to the apparent isotropic
obliquity distribution in Hot-Jupiter systems when taking tidal
realignment into account (i.e., omitting systems with relatively
strong tidal interactions).

Our results support the idea that the inward migration of close in
gas-giants is fundamentally different from the migration occurring in
compact multiple planet systems. It suggests that the planets we see
in the multiple planet systems might have migrated via disk-planet
interactions while Hot-Jupiters must have taken a different route.
Their path not only brought them close to their host stars but also
transported them out of the orbital plane of the disk out in which
they have formed.

\acknowledgments The authors warmly thank Guillaume H\'{e}brard for
supporting the observing campaign of KOI-94.  The authors are grateful
to Bill Cochran, Debra Fischer and Amaury Triaud for helpful
discussions, to Lauren Weiss for sharing an early version of her
group's manuscript about KOI-94, to John Brewer for pointing out an
error in an earlier version of this manuscript, and the {\it
  Kepler} team for creating an extraordinary tool for
discovery. Particular thanks are due to the {\it Kepler} follow-up
observing program for organizing the characterization of the host
stars. Work by S.A., J.N.W., and J.A.J.\ was supported by NASA Origins
award NNX09AB33G and the work by S.A.\ and J.N.W.\ was supported by
NSF grant no.\ 1108595. This research has made use of the following
web resources: {\tt simbad.u-strasbg.fr,
  adswww.harvard.edu,arxiv.org}.  The W.\ M.\ Keck Observatory is
operated as a scientific partnership among the California Institute of
Technology, the University of California, and the National Aeronautics
and Space Administration, and was made possible by the generous
financial support of the W.\ M.\ Keck Foundation. We extend special
thanks to those of Hawaiian ancestry on whose sacred mountain of Mauna
Kea we are privileged to be guests.

\end{document}